\documentclass[twocolumn,trackchanges]{aastex7}

\newcommand{\kms}{km~s$^{-1}$}

\usepackage{multirow}
\usepackage{amsmath}
\usepackage{rotating}
\usepackage{float}
\usepackage{comment}
\usepackage{adjustbox}
\usepackage{bm}

\begin{document}

\title{Frequent X-class Far UV Flares from a Young Solar Analog DS Tucanae A}

\author[0009-0006-0318-3385]{Tristen M. Sextro}
\affiliation{Pennsylvania State University, 201 Old Main, University Park, PA 16802, USA}
\email{tms7352@psu.edu}

\author[0000-0002-1002-3674]{Kevin France}
\affiliation{Laboratory for Atmospheric and Space Physics, University of Colorado Boulder, Boulder, CO 80303}
\affiliation{Department of Astrophysical and Planetary Sciences, University of Colorado Boulder, Boulder, CO 80309}
\affiliation{Center for Astrophysics and Space Astronomy, University of Colorado Boulder, Boulder, CO 80309}
\email{kevin.france@colorado.edu}

\author[0000-0003-4452-0588]{Vladimir S. Airapetian}
\affiliation{Sellers Exoplanet Environments Collaboration, NASA Goddard Space Flight Center, 8800 Greenbelt Rd, Greenbelt, MD, United States}
\email{vladimir.airapetian@nasa.gov}
\affiliation{Department of Physics, American University, Washington, DC, USA}

\author[0009-0000-6796-4056]{Grace M. Sweetak}
\affiliation{Lehigh University, 27 Memorial Dr W, Bethlehem, PA, United States}
\affiliation{NASA Goddard Space Flight Center, 8800 Greenbelt Rd, Greenbelt, MD, United States}
\affiliation{Pennsylvania State University, 201 Old Main, University Park, PA 16802, USA}
\email{gms423@lehigh.edu}

\author[0000-0002-3719-8212]{Randall L. McEntaffer}
\affiliation{Pennsylvania State University, 201 Old Main, University Park, PA 16802, USA}
\email{rlm90@psu.edu}

\author[0000-0002-1297-9485]{Kosuke Namekata}
\affil{Heliophysics Science Division, NASA Goddard Space Flight Center, 8800 Greenbelt Road, Greenbelt, MD 20771, USA}
\affiliation{The Catholic University of America, 620 Michigan Avenue, N.E. Washington, DC 20064, USA}
\affiliation{The Hakubi Center for Advanced Research, Kyoto University, Yoshida-Honmachi, Sakyo-ku, Kyoto 606-8501, Japan}
\affiliation{Department of Physics, Kyoto University, Kitashirakawa-Oiwake-cho, Sakyo-ku, Kyoto, 606-8502, Japan}
\email{namekata@kusastro.kyoto-u.ac.jp}

\author[0000-0002-0412-0849]{Yuta Notsu}
\affiliation{Laboratory for Atmospheric and Space Physics, University of Colorado Boulder, Boulder, CO 80303}
\affil{National Solar Observatory, 3665 Discovery Drive, Boulder, CO 80303, USA}
\affiliation{Department of Astrophysical and Planetary Sciences, University of Colorado Boulder, Boulder, CO 80309}
\email{Yuta.Notsu@colorado.edu}


\begin{abstract}
The young solar analog DS Tuc A (${\sim}45$ million years old, spectral type G6V) serves as an excellent proxy to the Sun during early stages of the Earth’s evolution. The young Sun's enhanced magnetic activity in the form of X-ray and extreme UV emission, stellar wind, and flares likely affected the magnetospheric and atmospheric environments of early Venus, Earth and Mars. Studies of DS Tuc A help constrain evolution of magnetospheric and atmospheric environments of early Venus, Earth, and Mars and create opportunities to study the impact of flares on the transiting exoplanet, DS Tuc Ab. We analyze ${\sim}11$ hours of Far-ultraviolet (FUV; 1130~--~1450~\AA) observations taken by the \textit{Hubble Space Telescope} Cosmic Origins Spectrograph during late 2023 and early 2024. We find DS Tuc A exhibited 26 flares over $\sim{11.1}$\,hours of observations in the transition region lines Si IV (1394\,\AA{} + 1402\,\AA{}), with durations ranging 90\,--\,960 seconds and FUV absolute energies ranging $\sim$~$10^{29}$\,--\,$10^{31}$ erg over the observed bandpass. Si IV flares are accompanied by increases of other chromospheric and transition region lines, however the strongest Si IV flares show no enhanced coronal emission traced by Fe XXI 1354\,\AA{}. We present an analysis of each detected flare, including X-ray and bolometric luminosities scaled from FUV flare energies, electron density measurements for the strongest flares, and the Far-UV continuum response to flaring. These results provide constraints on the frequency and energetics of flares comparable to X100-class solar flares and FUV-emitting regions in young solar analogs.

\end{abstract}

\keywords{}

\section{Introduction}
Young solar analogs in their first 600 Myr represent rapidly rotating stars with large convective envelopes that generate strong surface magnetic fields up to a few kG. The magnetic fields drive X-ray bright coronae, fast and massive stellar winds, and frequent energetic flares with energies up to 10$^{35}$ erg. Recent observations suggest that large stellar flares on solar-like stars are frequently associated with coronal mass ejections (CMEs) observed via Doppler shifts of H-${\alpha}$ (10,000K) and other warm ($\sim$ 100,000K) emission lines, X-ray dimming and Type II radio burst events \citep{2022NatAs...6..241N, namekata_2024a, namekata_2026,  
callingham_typeII_radio_burst}. X-ray and associated extreme UV fluxes from quiescent and flaring corona complemented by fast and massive CMEs are the sources of space weather that can significantly affect dynamics and chemistry of planetary atmospheres and their habitability \citep{prebio_chem_airapetian, spaceweather_airapetian}.

To characterize these impacts, we need to specify the frequency and energy of stellar flares from young solar analogs. Observations from missions such as \textit{Kepler} and \textit{TESS} have revealed that young G-type stars including DS Tuc A produce more energetic and frequent flares than those observed on the current Sun \citep{superflares_solaranalogs_maehara, spaceweather_airapetian, youngstar_superflares, statproperties_superflares, TESS_FFD}. These stellar flares are driven by intense magnetic activity, which is amplified in young stars due to strong surface magnetic fields powered by rapid rotation and deep convection zones. These events are often accompanied by fast and massive coronal mass ejections (CMEs), which eject vast quantities of magnetized plasma into space, potentially impacting surrounding planetary systems \citep{hu_CMEs_solarlikestars, 2022NatAs...6..241N, namekata_2024a, namekata_2026}.
The impact of these flares and CMEs on orbiting planets can be profound, especially for planets in the habitable zone around G-M dwarfs. High-energy radiation and energetic particles from flares can erode planetary atmospheres and alter its chemistry \citep{airapetian_2017, spaceweather_airapetian, chen_rocky_habzone_planet_chemistry, prebio_chem_kobayashi}. CMEs can compress planetary magnetospheres, induce geomagnetic storms, and enhance atmospheric escape processes \citep{spaceweather_airapetian}. For young Earth-like planets, this could mean significant loss of water and other volatiles, potentially compromising habitability. The cumulative effect of frequent flaring and CME activity during the early stages of stellar evolution is a critical factor in determining the long-term atmospheric retention and habitability of exoplanets. \\
Recent modeling studies, observations, and lab experiments suggest that the young Sun may have exhibited similar behavior, which has important implications for the evolution of Earth's atmosphere and the origin of life. Studies by \cite{prebio_chem_airapetian, prebio_chem_kobayashi} suggested that energetic particles from young solar flares could have driven prebiotic chemistry, contributing to the synthesis of essential biomolecules. However, for exoplanets orbiting more active stars, the balance between atmospheric erosion and chemical enrichment remains a key question. Understanding the frequency, intensity, and particle environment of stellar flares and CMEs from young solar analogs is thus essential for assessing the habitability of the planets around them.

DS Tucanae (DS Tuc) is a visual binary star system \citep{torres1} with a projected angular separation of 5.4" \citep{benatti} at a distance of ${\sim}44$ pc from the Earth \citep{gaia}. The system consists of a K3V star (DS Tuc B) and a G6V star (DS Tuc A) \citep{torres2}. DS Tuc A is the larger and hotter of the two stars in the system with a mass of 1.01$\pm$0.06 M$_\odot$, a radius of 0.964$\pm$0.029 R$_\odot$, and an effective temperature of 5428$\pm$80 K \citep{newton}. DS Tuc A has an age of $45\pm4$ million years placing it in the very early stages of its main sequence lifetime \citep{newton}. DS Tuc A also has one confirmed exoplanet, DS Tuc A b (M$<$1.3M$_\text{Jup}$, R=0.50$\pm0.02$R$_{\text{Jup}}$) \citep{benatti}. We consider DS Tuc A to be solar analog due to its size and spectral classification being so close to that of the Sun (G2V, T$_{\text{eff }}{\approx}$ 5778), thus findings about DS Tuc A can be applied to the Sun during its early stages of development. Through these findings, we can draw implications on the Sun's impact on terrestrial planets like Venus, Earth, and Mars in the early years.

The \textit{Hubble Space Telescope (HST)} provides high-resolution ultraviolet spectroscopy above the Earth's atmosphere, enabling precise measurements of stellar activity and variability. Its Cosmic Origins Spectrograph (COS; \citealt{cos}) delivers the highest UV throughput of any space-based spectrograph to data, providing time resolved, broad wavelength coverage ideal for detecting and characterizing flares in young solar analogs. The Primary Science Aperture (PSA) is 2.5" in diameter \citep{cos_handbook}, thus we are confident that the companion, located 5.4" from the primary, does not contaminate the COS observations.

In this paper, we present observations of DS Tuc A using the \textit{HST} COS to characterize the young solar analog's flaring state versus quiescence as well as to analyze each individual flare detected from the source throughout the observation. In Section \ref{sec:data_reduction} we describe the observations taken and the basic reduction method used. In Sections \ref{sec:flare_freq_analysis} and \ref{sec:flare_spec_analysis} we discuss both timing and spectral analysis of flares in the form of flare frequency and emission line fitting during flaring and quiescence.

\section{Data Reduction}
\label{sec:data_reduction}

\subsection{Observations}
DS Tuc A was observed by \textit{HST} in November 2023 and in March 2024 using the Cosmic Origins Spectrograph (COS) with the FUV G130M grating centered at 1291\,\AA{}. Eight 3-orbit visits were planned however two out of four planned visits in November failed as well as one orbit in the final visit in the March observations. The November observations consist of two 7046\,s observations and the March observations consist of three 7046\,s observations and one 4560\,s observation totaling nearly 40 ks exposure time. The FUV G130M grating captures a wavelength range from approximately 1130\,\AA{} to 1450\,\AA{} with a ${\sim}$14.5\,\AA{} gap in collected data between approximately 1274\,\AA{} and 1287.5\,\AA{}. DS Tuc A observation information is summarized in Table 1. The quiescent spectrum was used to estimate the EUV flux from DS Tuc A and is published in~\citet{france25}. 

\begin{table}[h]
    \centering
    \setlength{\tabcolsep}{1.5pt}
    \caption{DS Tuc A Observations}
    \begin{tabular}{c|c c c}
         Dataset & Date & Start [MJD] & Exposure [s] \\
         \hline \hline
         LF4201010 & 14 November 2023 & 60262.992   & 7045.568 \\
         LF4204010 & 18 November 2023 & 60266.155   & 7045.568 \\
         LF42Z1010 & 26 March 2024 & 60395.398   & 7045.568 \\
         LF42Z2010 & 27 March 2024 & 60396.123   & 7045.568 \\
         LF42Z3010 & 29 March 2024 & 60398.295   & 7045.536 \\
         LF42Z4010 & 30 March 2024 & 60399.085   & 4560.384 \\
    \end{tabular}
    \label{tab:observations}
\end{table}

\subsection{Reduction}

\textit{HST} data were taken in \texttt{TIME\_TAG} mode which allows whole exposures to split into smaller time bins for precise timing analysis. We used the \textit{HST} COS data reduction tools, COSTools, for splitting the data into various time cadences and the \textit{HST} COS data reduction pipeline, \texttt{calcos}, for data calibration. To split the data into different time cadences, we used the \texttt{COSTools} function \texttt{splittag} which takes processed \texttt{corrtag} files and a time increment. We began with a brief investigation on which time cadence works best as a balance between high temporal resolution and high signal to noise ratio (SNR). We considered four different time cadences including 500\,s, 100\,s, 60\,s, and 30\,s. 500\,s and 100\,s yielded insufficient temporal resolution such that most flares would not be visible in the light curves. 60\,s and 30\,s both did well with SNR but the 30\,s temporal resolution was most satisfactory thus we chose this time cadence for all light curve generation. The high temporal resolution is important as it is equivalent to the number of data points present in generated light curves which will impact measured flare fluxes and durations. The 100\,s and 500\,s cadence data served better for analysis of weaker lines such as the 1401.16\,\AA{} O IV line. The \texttt{splittag} function generated new \texttt{corrtag} files which were given to the \texttt{calcos} calibration pipeline to generate one dimensional spectra. The \texttt{calcos} pipeline was used with all default parameters and generated 1333 processed 30\,s exposure spectra (236 for the first five datasets, 153 for the sixth) for analysis. 

\subsection{Light Curve Generation}
\begin{figure*}[t]
  \centering
  \includegraphics[width=\textwidth]{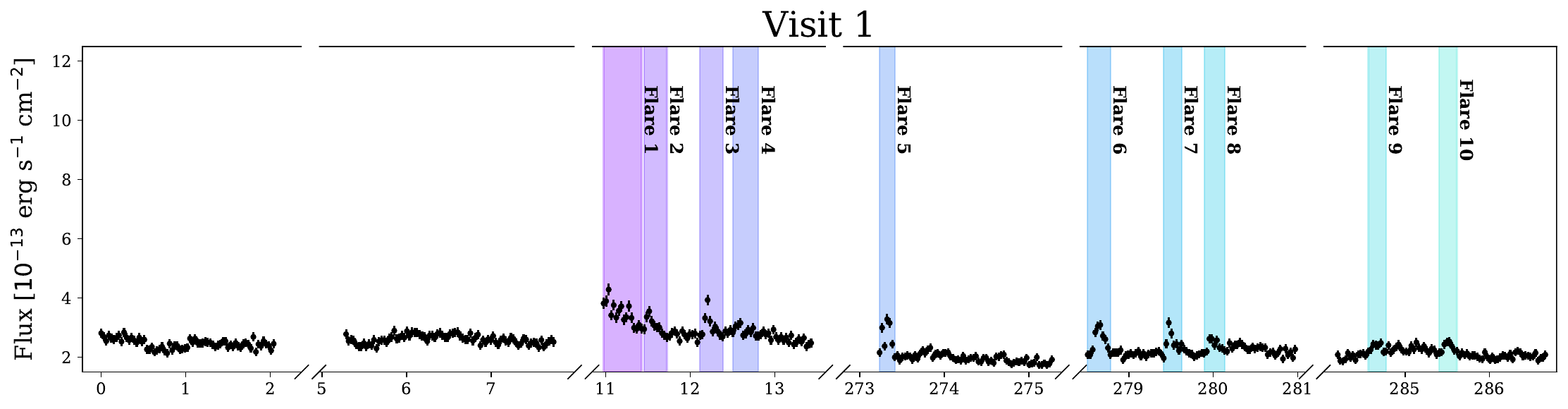}
  \includegraphics[width=\textwidth]{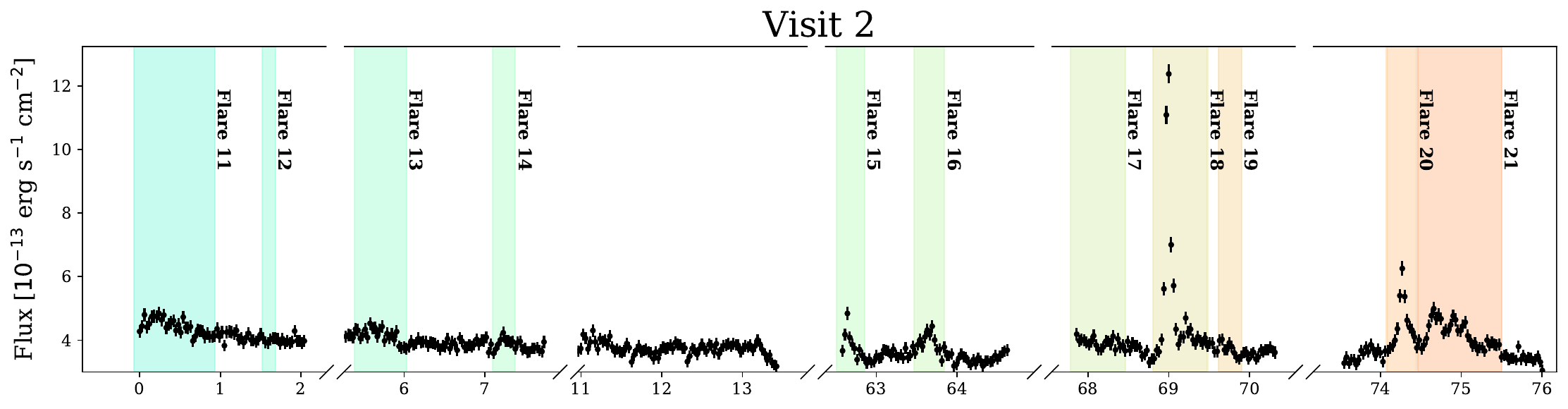}
  \includegraphics[width=\textwidth]{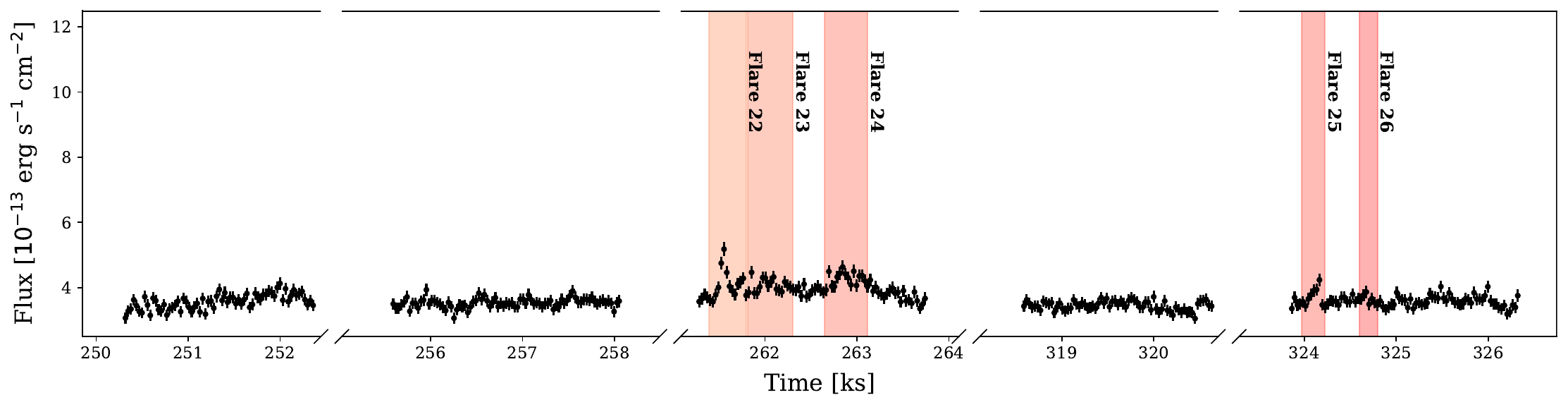}
  \caption{Light Curve taken over entire observation over entire wavelength band. Visit 1 refers to the November observations and Visit 2 refers to the March observations. Airglow lines including Ly$\alpha$ 1216\,\AA{}, NI (1200\,\AA{}), and the OI triplet (1302\,\AA{}, 1305\,\AA{}, and 1306\,\AA) have been excluded from this light curve. Time depicts the amount of time that has passed since the beginning of the visit with the top panel being the first visit and the bottom two panels being the second visit. Light curve timing cadence here is 30 seconds}
  \label{fig:white-light_lc}
\end{figure*}

Light curve flux values were generated by numerically integrating over a particular wavelength range determined by the emission line of interest. The exposure end time for each spectrum was taken as the time coordinate of the flux data point. Lines of interest are denoted in Table \ref{tab:linesOfInterest} with their rest wavelength \citep{NIST_ASD} and the region of the stellar atmosphere they probe \citep{lines_regions_peter}. To generate the light curve seen in Figure \ref{fig:white-light_lc}, the entire observed bandpass is integrated over with the exception of several airglow spectral lines which were simply cut out of the data. The airglow lines are spectral lines seen in the data as a result of the Earth's atmosphere coming between the \textit{HST} and the target thus emission is seen from the most prominent ions in the Earth's atmosphere. These lines include Ly-$\alpha$ ($\lambda$\,1215\,\AA{}), the OI triplet ($\lambda \lambda \lambda\,1302$\,\AA{}, 1305\,\AA{}, 1306\,\AA{}), and NI ($\lambda\,1200$\,\AA{}). For the purpose of this work, no reconstruction is done for any of the above listed airglow lines as they are simply masked out of all generated light curves.

\begin{table}[]
    \centering
    \small
    \setlength{\tabcolsep}{2pt}
    \caption{FUV Emission Lines of Interest. TR = Transition Region}
    \begin{tabular}{l|p{30mm}|c|l}
       Ion    & $\lambda_{\text{0}}$ [\AA\ ] & log T [K] & Origin Region \\
       \hline \hline
       C II   & 1334.532, 1335.708 & 4.4 & Chromosphere \\
       Si III & 1206.510       & 4.7 & TR \\
       C III  & 1174.933, 1175.263, \newline 1175.590, 1175.711, \newline 1175.987, 1176.370      & 4.8 & TR \\
       Si IV  & 1393.744, 1402.770 & 4.9 & TR \\
       O IV   & 1401.156       & 5.2 &    TR   \\
       N V    & 1238.821, 1242.804 & 5.3 & TR \\
       O V    & 1371.292       & 5.4 & TR \\
       Fe XXI & 1354.080       & 7.1 & Corona \\
       
    \end{tabular}
    \label{tab:linesOfInterest}
\end{table}

\subsection{The Hunt for Flares}

As a result of our small dataset, we adopt a straightforward ``by-eye" approach to identifying flares. We follow an approach similar to that in \citet{feinstein_paper} where we find large amplitude outliers in the light curve which are followed by a decay. Following this method, we ensure at least two data points are above the base noise level of the light curve for each individual continuous observation. For identifying flares, we looked at the light curve taken over only the Si IV emission line at 1393.76/1402.77\,\AA{} where we detected 26 flares. In addition, we analyzed light curves from Si III (1206\,\AA{}), the C III multiplet (1175\,\AA{}), and C II (1335/1336\,\AA{}), however no additional flares were found. Figure \ref{fig:white-light_lc} depicts the light curve taken over the entire FUV bandpass (with airglow emission lines removed) with flares detected from the Si IV  light curve marked.

\subsection{Generating Quiet/Flaring Spectra}
\label{subsection:generating quiet/flare spectra}
Flaring spectra are created by adding all of the spectra together from light curve points within a specific flare, then dividing by the number of spectra. The master quiescent spectrum is created by then adding together all of the light curve points that were not present in a flare then dividing by the number of spectra. Spectra are added together by linearly adding each flux value and calculating uncertainties directly using the total number of counts with equations for upper and lower error bounds from \cite{stats}. All spectra from \texttt{splittag} do not come on a single, uniform wavelength array so prior to addition, we create a wavelength array with 0.01\AA\ spacing as this is the approximate spacing on the original data. All flux and error values are interpolated on this wavelength array. In addition, we notice a systematic wavelength shift between the first and second portions of observation of approximately 0.135\,\AA{} corresponding to about 28\,km\,s$^{-1}$ at 1450\,\AA{} and 36\,km\,s$^{-1}$ at 1130\AA{}\,. This appears to be related to differences in target acquisition between the November and March observations. To correct for this, data from the March observations are shifted onto the same wavelength grid as data from the November observations such that the 1206\,\AA{} Si III emission lines match up. This Si III line is chosen as it is one of the strongest single lines seen in the data.

\subsection{Identifying and Fitting Emission Lines; Creating Line-List}
\label{section: identifying lines/line list}

From the generation of master quiescent and flaring spectra described in Section \ref{subsection:generating quiet/flare spectra}, emission lines were identified and fit using Python's \texttt{lmfit}  \cite{lmfit}. Initial lines were identified based on previous studies of emission line profiles highlighting a range of formation temperatures \cite{feinstein_paper} \cite{pagano_paper}. Previous studies have fit identified emission lines with a single gaussian profile and double-gaussian profiles \cite{feinstein_paper, pagano_paper}. To compare differences between various models, we fit 5 different models on these initial ions, including Gaussian (single and double), Lorentzian, Voigt and Pseudovoigt.

Models were constructed with  \texttt{lmfit} by constructing a built-in model, convolving with the COS line spread function (LSF), and interpolating to create a new model. This process creates an \texttt{lmfit} model with the same parameters and definitions of the built-in modes, but accounts for the light distribution at the focal plane by including the COS LSF. In addition, we tested each built in model by adding a linear component to represent the emission continuum. However, we found that the linear component coinciding with a peaked profile constrained our model too much and did not allow for error extraction. Similarly, using the double-gaussian profile constrained the model, leading to reduced $\chi^{2}$ values $\ll$ 1. Ultimately, we decided to fit all ions with a Pseudovoigt profile, a weighted sum of a Gaussian and Lorentzian profile. A Pseudovoigt model resulted in stochastic residuals, and allowed us to extract both error values and information regarding the composition of the peak, in other words, whether it resembled a more Gaussian or Lorentzian profile. 

After choosing a fitting profile and fitting initial lines of interest, we carefully inspected the master quiescent spectrum to identify weaker emission lines. In some cases, emission lines were identified extremely close together, in which fitting with a single peak was not reasonable. In these cases, we fit with a number of Pseudovoigt models depending on the lines of interest. 

\begin{figure}[h!]
  \centering
  \begin{tabular}{@{}c@{}}
    \includegraphics[width=\linewidth,height=150pt]{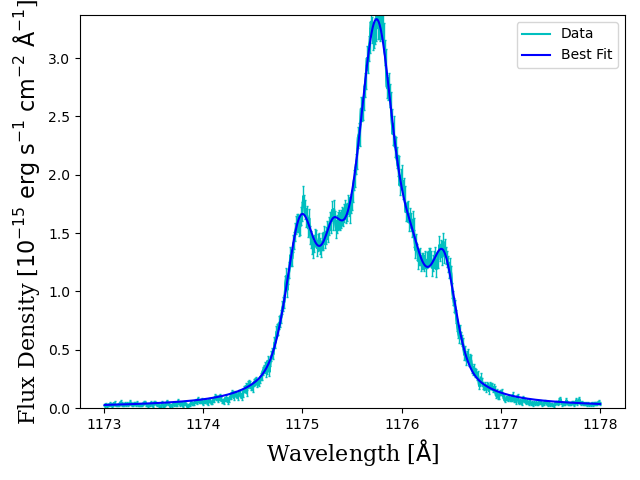} \\[\abovecaptionskip]
    \small (a)  
  \end{tabular}

  \vspace{\floatsep}

  \begin{tabular}{@{}c@{}}
    \includegraphics[width=\linewidth,height=150pt]{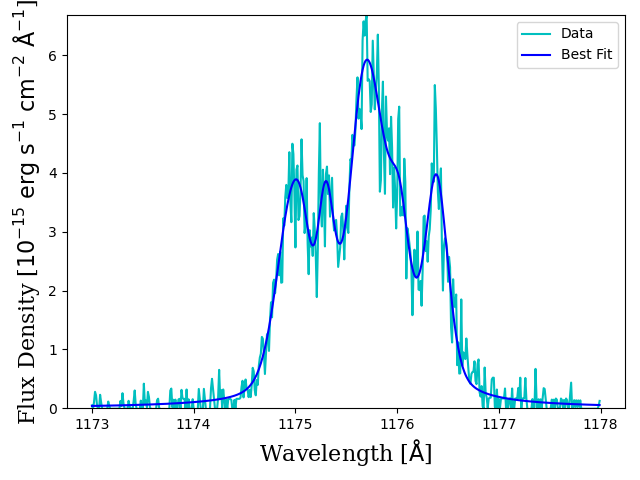} \\[\abovecaptionskip]
    \small (b)  
  \end{tabular}

  \caption{C III multiplet fit with 6 separate pseudovoigt profiles added linearly, for both the quiescent spectrum (a) and flare 18 (b). }
  \label{fig:C III multiplet}
\end{figure}

An example of one of these cases was C III between 1174 - 1178\,\AA. The multiplet contains six identifiable C III transitions. Due to the strength of the lines, we were able to fit the quiescent spectrum and most lines in the individual flaring spectrum. Figure \ref{fig:C III multiplet} depicts the fit of this C III multiplet for the quiescent spectrum and the flare 18 spectrum.

The 6 identifiable lines were fit with a model consisting of six individual pseudovoigt profiles added linearly. Each profile was initially centrally constrained by their laboratory wavelengths. The two of the six peaks central to the multiplet were initially constrained in their amplitude by the maximum flux of that section, while the other four peaks were constrained by the mean amplitude between 1174\,\AA{} and 1178\,\AA{}. Following the same process, either with a single peak or a multiplet, all identifiable ions were fit and their parameters are listed in Table \ref{tab:line_catalog}.

We identified lines in comparison to the NIST Atomic Spectra Database. Table \ref{tab:line_catalog} lists all identified lines and their extracted model values including laboratory wavelength, observed wavelength, for the master quiescent spectrum and Flare 18, one of the strongest flares identified. Observed wavelength and velocity shift are determined using the extracted model values from the master quiescent spectrum. It is important to note that NIST occasionally did not list an uncertainty associated with the laboratory wavelength, therefore, velocity shift uncertainty could not be calculated in these cases. Furthermore, the profile fitting Flare 18 with the ion C III contained six separate pseudovoigt profiles, and due to the lack of light curves present for Flare 18 compared to the master quiescent spectra, uncertainty could not be extracted for Flare 18 between 1174\,\AA{} and 1178\,\AA{}. 

\section{Flare Frequency Analysis}
\label{sec:flare_freq_analysis}

As can be seen in Figure \ref{fig:white-light_lc}, we detect 26 distinct flares in the \textit{HST} COS data. For each ion in which significant flaring is detected, we create flare frequency distributions and fit them with power laws. Additional flare analysis is done in Section \ref{sec:flare_spec_analysis} consisting of line profile fitting for the larger flares, continuum fitting for each flare, measuring flare electron density from the Si IV/O IV emission line ratio, and a discussion on scaling the FUV flare energy to X-ray and bolometric energies. For clarity, FUV energy refers to energy calculated over the observed 1130~--~1450\,\AA{} bandpass.

\subsection{Flare Frequency Distribution}
\label{sec:ffds}
Flare frequency distributions (as described in \cite{loyd_paper} and henceforth referred to as FFDs) are cumulative distributions of the frequency of flares as a function of a parameter of the flares.  It is typical to plot the frequency of flares as a function of the flare's absolute energy (E) or equivalent duration ($\delta$). Absolute energy is simply the time integral taken over the flaring flux with the quiescent flux subtracted during all flaring time. This is then multiplied by the spherical area calculated using the distance to the source to convert from flux into energy. The equation used to describe the absolute energy is given in Equation \ref{eq:abs_ener}. 
\begin{equation}
    E = 4\pi d^2 \int_{\text{flare}}\left(F_{\mathrm{f}}-F_{\mathrm{q}}\right)\mathrm{dt}
    \label{eq:abs_ener}
\end{equation}
where F$_{\text{f}}$ represents the flux during flaring times, F$_{\text{q}}$ represents the flux during quiescence, and d depicts the distance to the source. For all luminosity and absolute energy calculations, a distance of $44.0622^{+0.0693}_{-0.0691}\,$pc from \textit{GAIA} DR2 \citep{gaia} is adopted. 

Equivalent duration is described in \cite{equiv_dur_paper} as the integral over the difference between flaring and quiescent flux normalized by the quiescent flux in the same spectral band. Equivalent duration can also be thought of as the temporal counterpart to the spectroscopic equivalent width \citep{loyd_paper}. Put more intuitively, equivalent duration can be thought of as the additional amount of time following the base flare duration it would take for the source in quiescence to emit the total flaring flux over the course of a flare. The equation used to describe equivalent duration is given in Equation \ref{eq:equiv_dur}.
\begin{equation}
    \delta = \int_{\text{flare}}\left(\frac{F_{\mathrm{f}}-F_{\mathrm{q}}}{F_{\mathrm{q}}}\right)\mathrm{dt}
    \label{eq:equiv_dur}
\end{equation}

Following a process similar to \cite{loyd_paper}, we create FFDs for the ions generating the largest flares in the source including Si IV ($\lambda \lambda 1394\text{\,\AA{}},1402\text{\,\AA{}}$) Si III ($\lambda 1206\text{\,\AA{}}$), C II ($\lambda \lambda 1334\text{\,\AA{}}, 1335\text{\,\AA{}}$), and C III ($\lambda 1176\text{\,\AA{}} $ multiplet) These FFDs are then fit to a power law as is seen in Equation \ref{eq:power_law}. 
\begin{equation}
    \nu = 
    \begin{cases}
        C \left( \frac{E}{10^{30}\,\mathrm{erg}} \right)^{-\Gamma} \\[8pt]
        C \left( \frac{\delta}{1\,\mathrm{s}} \right)^{-\Gamma}
    \end{cases}
    \label{eq:power_law}
\end{equation}
where $\nu$ is the cumulative flare rate in flares per day, $\Gamma$ are the power law indices, C are the proportionality constants for fitting purposes, and E and $\delta$ are the absolute energy and equivalent durations respectively. Steeper power laws, i.e. larger values of the positive index $\Gamma$ in C$\times$E$^{-\Gamma}$, imply that low-energy flares occur more frequently, whereas smaller $\Gamma$ (closer to zero), corresponding to a flatter, shallower distribution, imply relatively more frequent high-energy flares.

Noting that the FUV FFDs for DS Tuc A are derived from a small sample of flare (26), we find power law indices for fits to each of the FFDs in Figure \ref{fig:ffds} for Si IV, Si III, C III, and C II of $1.1\pm0.06$, $0.82\pm0.07$, $0.89\pm0.05$, and $0.91\pm0.09$ respectively for absolute energy and $1.2\pm0.06$, $0.81\pm0.07$, $0.90\pm0.05$, and $0.91\pm0.09$ respectively for equivalent duration. 

\begin{table}
    \centering
    \begin{tabular}{m{0.75cm} m{2cm} m{2cm} m{2cm} }
      Flare & E [$10^{30}$ erg]& $\delta$ [s]  & Duration [s] \\
        \hline \hline
   1 & $    9.77 \pm 0.29$ & $   165.9 \pm    21.05 $ & $ 450 $ \\
   2 & $    3.47 \pm 0.23$ & $    59.0 \pm     5.82 $ & $ 270 $ \\
   3 & $    3.17 \pm 0.23$ & $    53.8 \pm     5.31 $ & $ 270 $ \\
   4 & $    2.69 \pm 0.24$ & $    45.6 \pm     4.76 $ & $ 300 $ \\
   5 & $    0.77 \pm 0.19$ & $    13.1 \pm     1.05 $ & $ 180 $ \\
   6 & $    0.14 \pm 0.23$ & $     2.4 \pm     0.23 $ & $ 270 $ \\
   7 & $    0.25 \pm 0.20$ & $     4.2 \pm     0.37 $ & $ 210 $ \\
   8 & $    0.84 \pm 0.22$ & $    14.3 \pm     1.33 $ & $ 240 $ \\
   9 & $    1.03 \pm 0.20$ & $    17.5 \pm     1.52 $ & $ 210 $ \\
  10 & $    5.78 \pm 0.20$ & $    70.7 \pm     6.15 $ & $ 210 $ \\
\hline
  11 & $    19.8 \pm 0.41$ & $   242.7 \pm    42.93 $ & $ 930 $ \\
  12 & $    0.96 \pm 0.13$ & $    11.8 \pm     0.67 $ & $ 90 $ \\
  13 & $    8.59 \pm 0.33$ & $   105.2 \pm    14.92 $ & $ 570 $ \\
  14 & $    2.26 \pm 0.20$ & $    27.6 \pm     2.40 $ & $ 210 $ \\
  15 & $    2.24 \pm 0.23$ & $    27.4 \pm     2.71 $ & $ 270 $ \\
  16 & $    3.27 \pm 0.24$ & $    40.1 \pm     4.18 $ & $ 300 $ \\
  17 & $    5.46 \pm 0.34$ & $    66.9 \pm     9.70 $ & $ 600 $ \\
  18 & $    22.7 \pm 0.34$ & $   278.1 \pm    40.35 $ & $ 600 $ \\
  19 & $    0.89 \pm 0.20$ & $    10.9 \pm     0.95 $ & $ 210 $ \\
  20 & $    7.89 \pm 0.24$ & $    96.6 \pm    10.07 $ & $ 300 $ \\
\hline
  21 & $    15.4 \pm 0.42$ & $   189.2 \pm    33.94 $ & $ 960 $ \\
  22 & $    5.18 \pm 0.25$ & $    63.4 \pm     6.93 $ & $ 330 $ \\
  23 & $    5.45 \pm 0.29$ & $    66.8 \pm     8.20 $ & $ 420 $ \\
  24 & $    7.11 \pm 0.28$ & $    87.0 \pm    10.32 $ & $ 390 $ \\
  25 & $    1.31 \pm 0.19$ & $    16.0 \pm     1.29 $ & $ 180 $ \\
  26 & $    0.42 \pm 0.15$ & $     5.2 \pm     0.34 $ & $ 120 $ \\
        
    \end{tabular}
    \caption{List of FUV flare parameters derived from Equations \ref{eq:abs_ener} and \ref{eq:equiv_dur} for absolute energy and equivalent duration respectively as well as total estimated flare duration from light curves. Flare durations are in multiples of 30 seconds due to the 30 second time cadence defining the beginning and ending of flare times.}
    \label{tab:flare_params}
\end{table}

\subsubsection{Comparison with the FUV FFDs of M Dwarfs}

\begin{figure*}[h]
    \centering

    \includegraphics[width=0.99\textwidth]{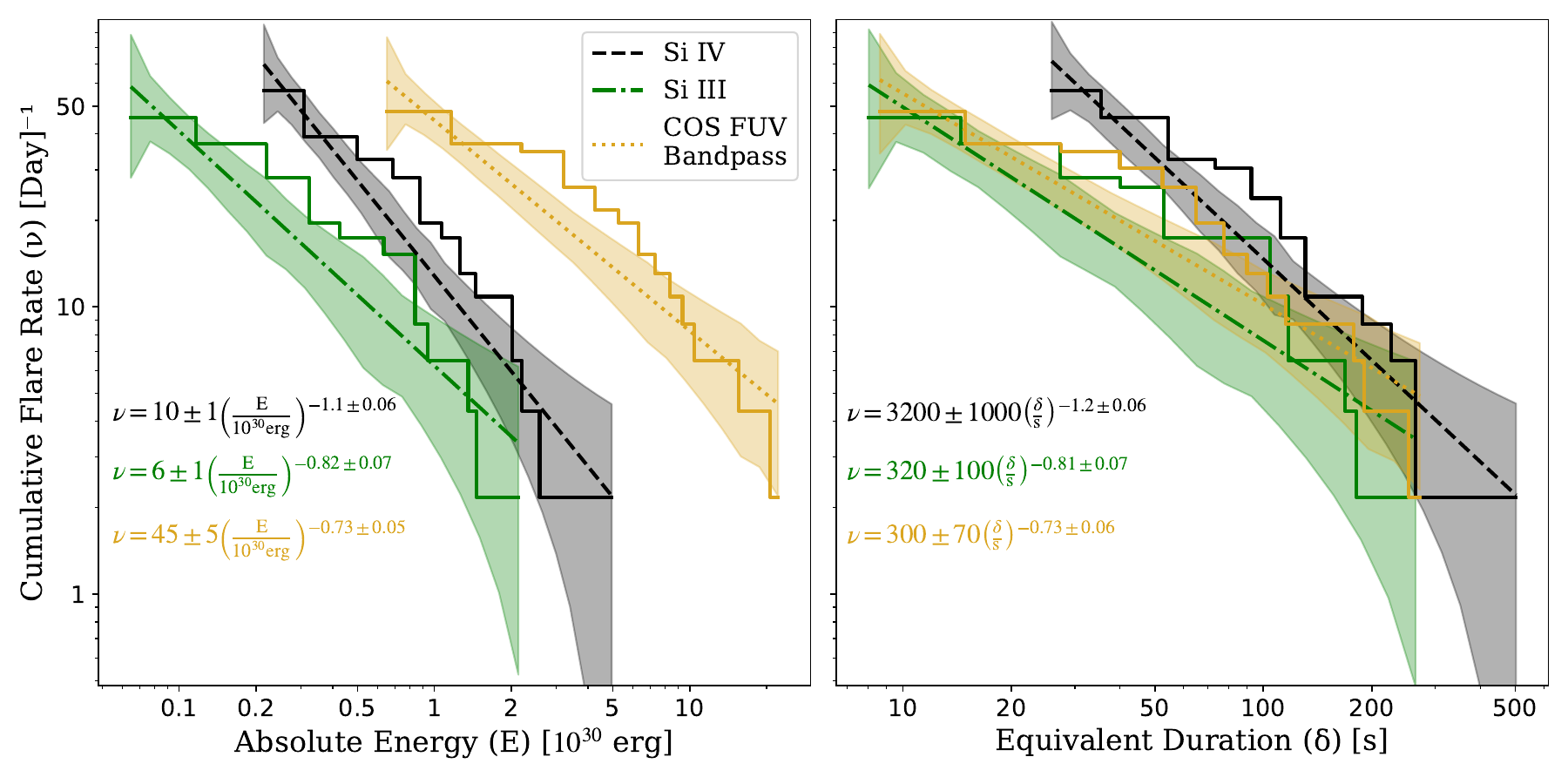}
    \includegraphics[width=0.99\textwidth]{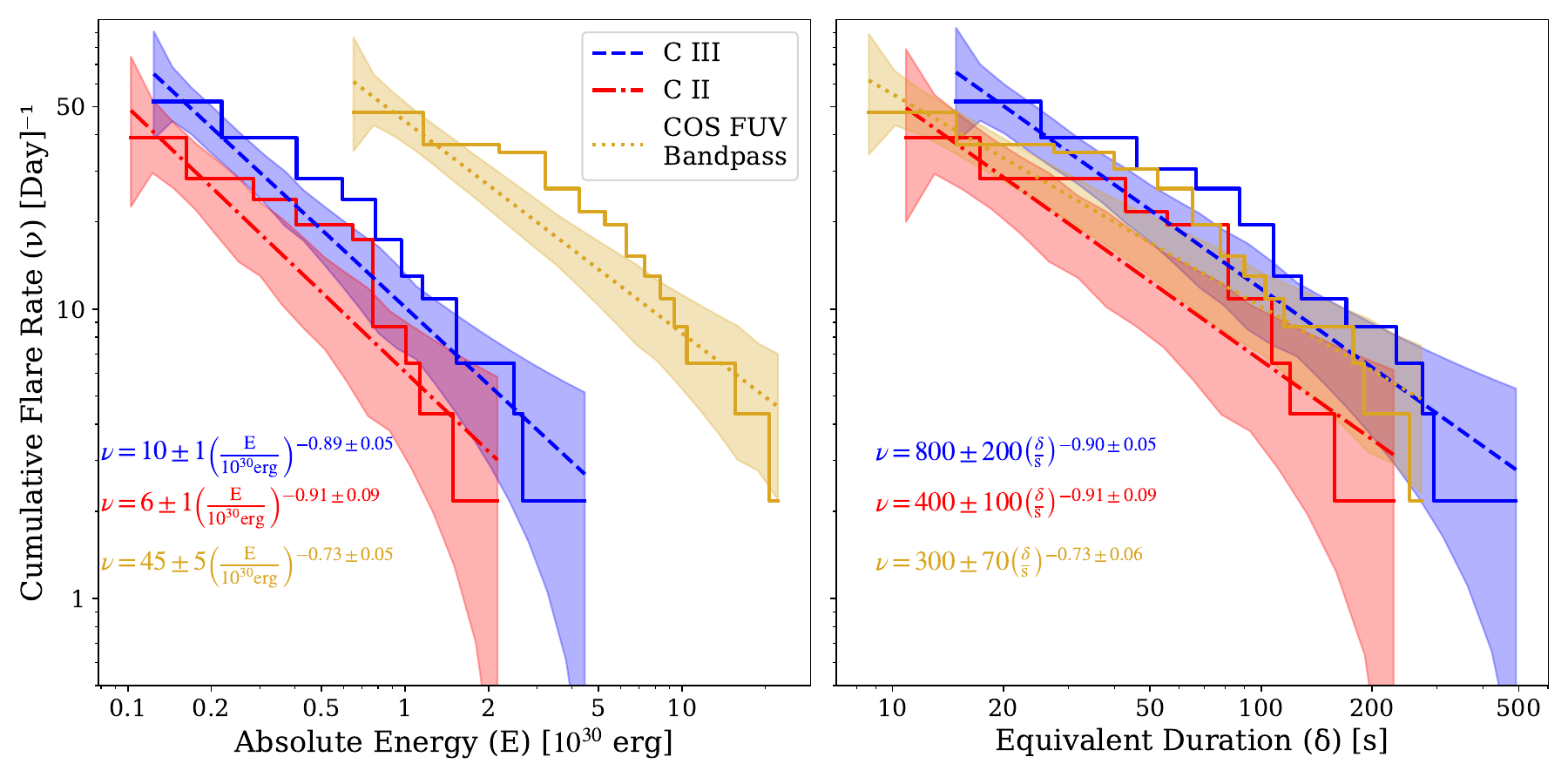}

    \caption{Flare frequency distributions for Silicon (top) and Carbon (bottom) lines with power-law fits overlaid. Both absolute energy (left) and equivalent duration (right) representations are shown. FFDs created over the observed FUV wavelength band are shown for reference titled `COS FUV Bandpass'. }
    \label{fig:ffds}
\end{figure*}

Work done by \cite{loyd_paper} analyzes FUV flares of 6 active and 4 inactive M dwarfs and creates FFDs in both absolute energy and equivalent duration similar to those in Figure \ref{fig:ffds}. Their work utilizes the COS G130M covering a bandpass of 1170\,\AA{}--1430\,\AA{}, with 1270\,\AA{}--1330\,\AA{} removed. We cover a similar FUV bandpass of 1130\,\AA{}--1430\,\AA{} with 1275\,\AA{}--1285\,\AA{} removed for this analysis. The flares seen in the active M dwarfs are much larger than those seen in DS Tuc with energies up to $10^{31}$ erg as well as equivalent durations reaching $10^4$ seconds. In equivalent duration space, \cite{loyd_paper} finds a power law index for the inactive M stars to be $0.77^{+0.15}_{-0.17}$ and $0.80^{+0.13}_{-0.14}$ for the active M stars. Over the COS FUV bandpass, we find a power law index of $-0.73\pm 0.05$ over both absolute energy and equivalent duration space. For both active and inactive M dwarfs, the FFD power law index is roughly consistent with that of DS Tuc. This indicates DS Tuc emits a similar number of high energy FUV flares per low energy flare compared to both active and inactive M dwarfs.


\subsubsection{Optical FFDs with \textit{TESS}}

Observations of DS Tuc by the \textit{Transiting Exoplanet Survey Satellite} (\textit{TESS}) and an analysis of flares seen in the \textit{TESS} bandpass is done in \cite{TESS_FFD}. The \textit{TESS} observations analyzed consist of three 2-minute cadence, 25-day light curves, sufficient for a thorough analysis of individual flaring events as well as to estimate flare frequencies \citep{TESS_FFD}. The work generates a flare frequency distribution for DS Tuc and fits it with a cutoff power law with its cutoff at $10^{34}$ erg \citep{TESS_FFD}. The cutoff power law for the \textit{TESS} FFD yields a power law index of $0.90\pm0.17$ \citep{TESS_FFD}. This power law index is consistent with several of those calculated for the \textit{HST} FUV data indicating DS Tuc emits FUV flares following a similar energy distribution as the optical flares detected by \textit{TESS}. 

\subsubsection{X-Ray Flares with XMM-Newton and Chandra}

Work done by \cite{xmm_flares} and \cite{chandra_flares} discuss the X-ray flaring behavior of DS Tuc, however each only find two strong flares from the source over 70\,ks and 57\,ks respectively. The \cite{xmm_flares} analysis of the work revealed the two flares released a total of $5-8\times10^{34}$\,erg over the 0.3 to 10.0\,keV X-ray bandpass as well as $0.9-2.7\times10^{33}$\,erg over the EUV band 200-300\,nm. Similarly, the \cite{chandra_flares} analysis derives flare energies of $8.6\times10^{33}$ erg and $3.7\times10^{33}$\,erg between the two detected flares across the 0.5 to 10.0\,keV X-ray bandpass. Unfortunately, it does not appear a conclusive FFD currently exists for DS Tuc in the X-ray as too few flares have been detected to create a statistically meaningful distribution.

\subsubsection{An FFD Interpretation}

The FUV flare frequency distributions for Si IV, Si III, C II, and C III are all well described by similar power-law indices in both absolute energy and equivalent duration, and the values agree within their respective uncertainties. This lack of a significant ion-to-ion variation suggests that the observed flares represent a single underlying population whose energy distribution is largely independent of the specific transition-region tracer used. In other words, while individual events differ in amplitude and duration, the statistics of how often DS Tuc produces flares of a given strength appear consistent across the major FUV ions, and with the FFD derived over the full COS bandpass. Given the limited sample of 26 flares and restricted dynamic range, these indices should be interpreted cautiously, but there is no evidence for strong wavelength- or formation temperature-dependent trends within the FUV.

\begin{figure}
    \centering
    \includegraphics[width=0.98\linewidth]{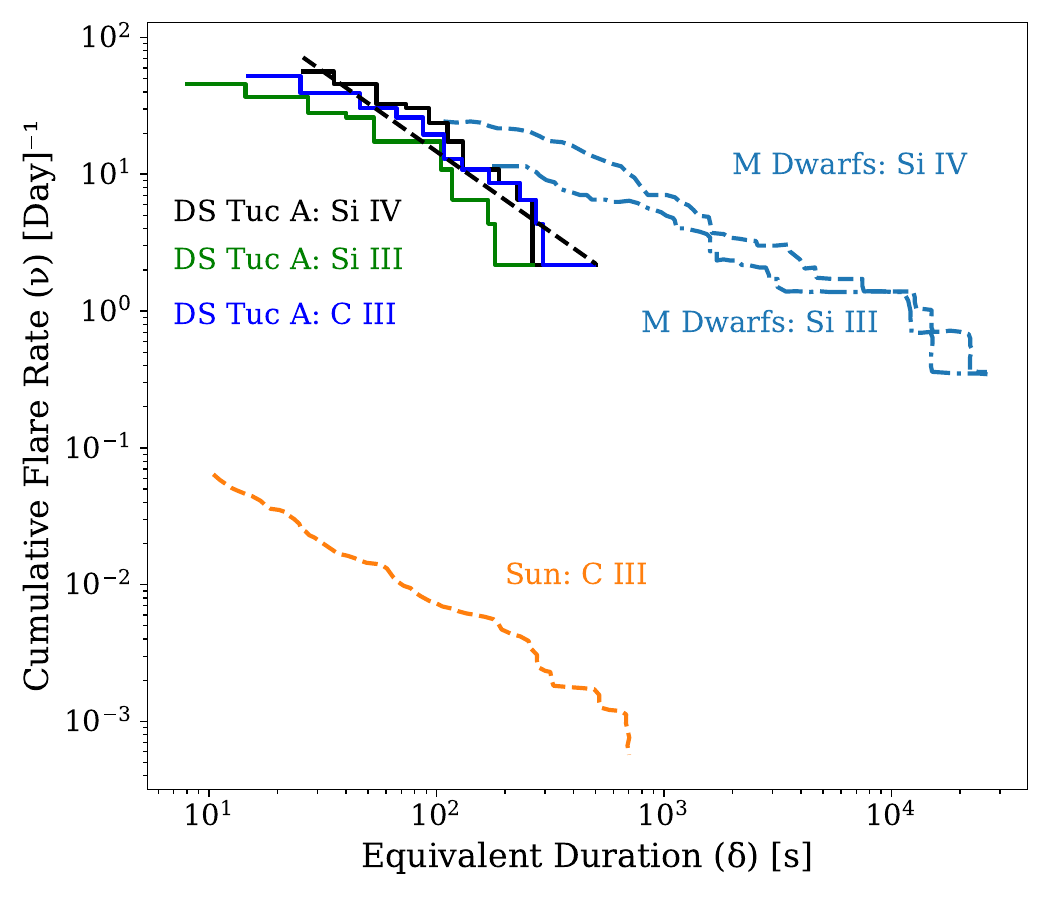}
    \caption{Comparison between DS Tuc A (solid black, blue, and green), M dwarfs (dashed blue), and the Sun (orange) as FFDs in equivalent duration space. Sun and M Dwarf data and distributions come from \cite{loyd_paper}.}
    \label{fig:comparison_ffd}
\end{figure}

In comparison to other bands, the FUV FFDs on DS Tuc are modestly consistent with those measured for both active and inactive M dwarfs in the FUV as well as the \textit{TESS} white-light FFD of the same star. This implies that, relative to low-energy events, DS Tuc roughly produces high-energy flares nearly as often in the COS FUV band as in the broader optical bandpass, as well as to typical M-dwarf flare stars at similar energies. 
See Figure \ref{fig:comparison_ffd} for a comparison DS Tuc A FFDs in Si IV, Si III, and C III with M dwarfs in Si IV/III and the Sun in C III. The figure is primarily Figure 8 from \cite{loyd_paper} with the relevant ions from M dwarfs and the Sun taken with DS Tuc A FFDs overlaid. This comparison helps to convey the similarity between DS Tuc A's flare distribution ($\Gamma=0.73\pm0.05$) and those seen in M dwarfs ($\Gamma=0.76^{+0.10}_{-0.09}$) and shows both DS Tuc A and M dwarfs are significantly more active than the modern day Sun by about 2-3 orders of magnitude at a given equivalent duration.
The sparse X-ray flare detections reported with XMM-Newton and Chandra indicate that DS Tuc is capable of producing large high-energy events, but the small number of X-ray flares prevents constructing a comparable FFD; thus, the most robust view of the flare statistics currently comes from the FUV and \textit{TESS} data.

\section{Flare Spectral Analysis and Interpretation}
\label{sec:flare_spec_analysis}

The flare spectral analysis will first discuss the path taken for finding the ideal emission line fitting function including models and fit statistics used. Following, we discuss fitting emission lines through specific flares to probe flaring plasma behavior relative to quiescence and  briefly discuss modeling the continuum to investigate the continuum response to flaring. We then discuss one by one each ion through Table \ref{tab:linesOfInterest} and how they respond through quiescent and flaring states. We discuss the Si IV / O IV density diagnostic to estimate the electron density in longer duration flares and conclude with a discussion on an FUV proxy for X-ray flare energy and scaling this to achieve bolometric flare energies.

\subsection{The Emission Line Model}
\label{subsubsection:emission line model}

\begin{figure}[h!]
    \centering
    \includegraphics[width=0.95\linewidth]{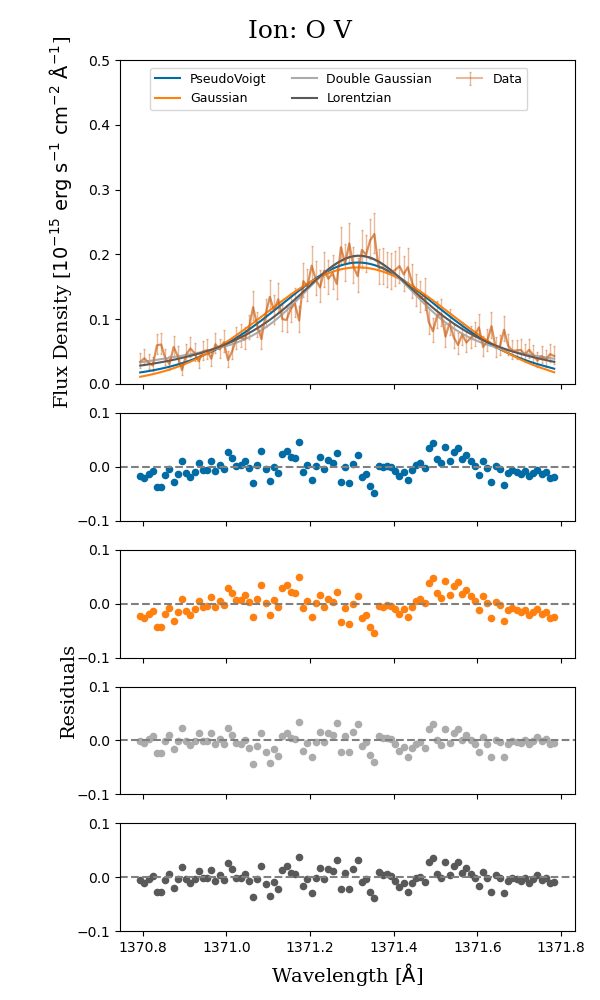}
    \caption{Example of the fits tested for O V centered at 1371.292\,\AA{}. Four fits tested overall, with models overlaying data and the corresponding residuals.}
    \label{fig:residuals}
\end{figure}

Mentioned in Section \ref{section: identifying lines/line list}, we decided to fit emission lines with a Pseudovoigt profile convolved with the COS LSF. While testing fit profiles, some lines fit better to a Gaussian, while some fit better to a Lorentzian, as measured by their reduced $\chi^{2}$ value and residual plot. Figure \ref{fig:residuals} depicts an example of this, showing the residual plot for O V. 

While all residuals are generally stochastic, reduced $\chi^{2}$ values vary between models. For the emission line in Figure \ref{fig:residuals}, the double gaussian model resulted in the lowest reduced $\chi^{2}$ value of 0.656. However, overfitting with the double gaussian model does not allow for error extraction of some parameters, such as flux. In this example, the reduced $\chi^{2}$ values for the Gaussian model and Lorentzian model were 1.515 and 0.701, respectively. This suggests that a Lorentzian model is the appropriate choice, however, some other emission lines are better represented by a Gaussian model. For example, the Fe XXI emission line centered at 1354.080\,\AA{} gives reduced $\chi^{2}$ values of 1.737 for a Lorentzian model, and 0.660 for a gaussian model. To account for both cases, we decided to use a Pseudovoigt profile, as it weighs both options. To further investigate effective fitting profiles, we also considered adding a linear component to the Pseudovoigt model to encompass the background noise. However, when including a linear component to the original built in Pseudovoigt profile, less than 50 percent of the fit statistics were able to produce an associated uncertainty. We believe this was due to over constraining the model, and ultimately decided to remove the linear component and fit with a Pseudovoigt profile only. This resulted in a  reduced $\chi^2$ of 0.66, once again suggesting slight overfitting but a more informative model, as every single peaked ion produced an associated uncertainty. 

\subsection{Emission Lines in Individual Flares}

\begin{figure*}[h]
    \centering
    \includegraphics[width=0.99\linewidth]{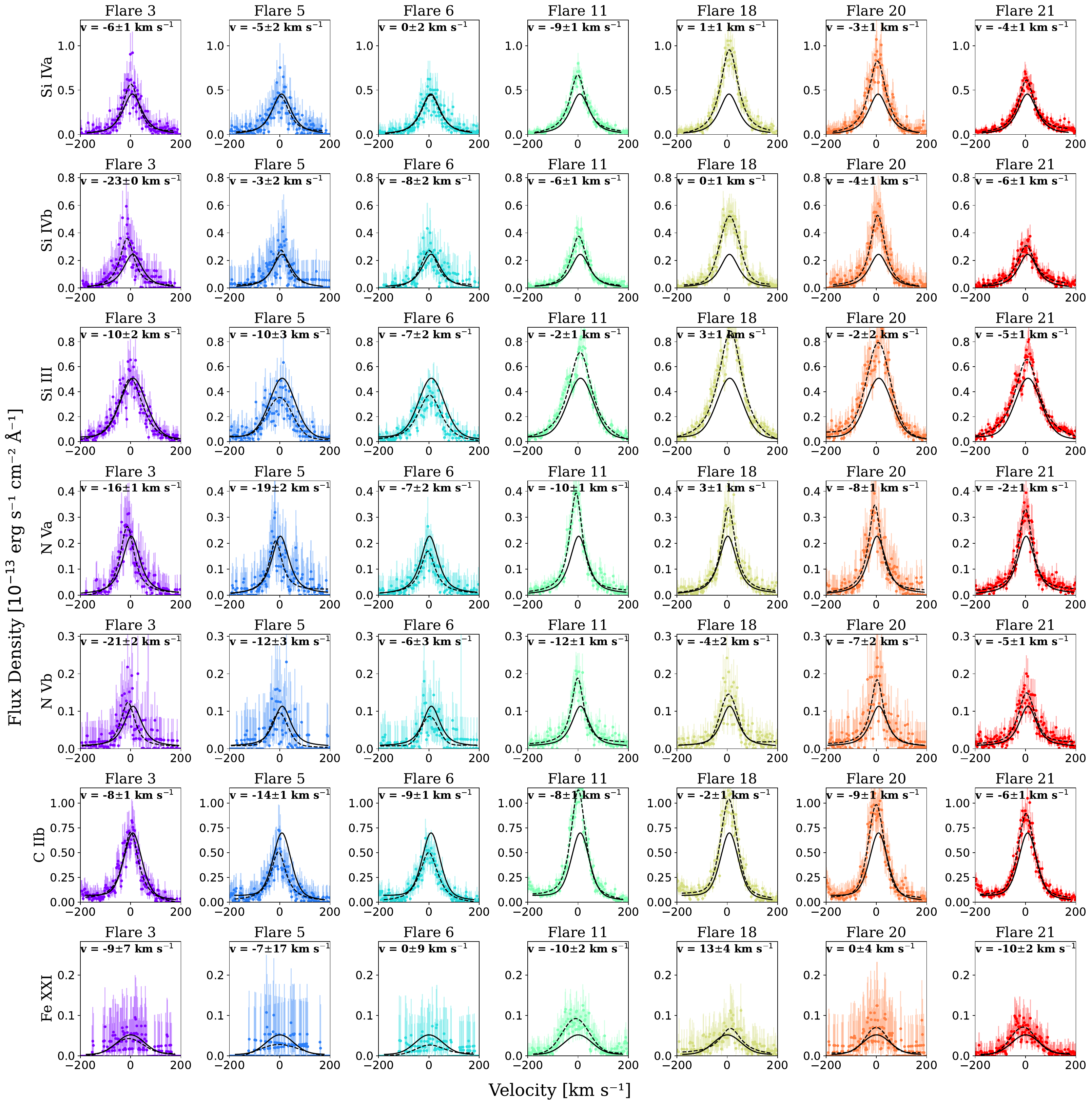}
    \caption{Best fit spectral line profiles for specific ions in chosen flares. Doublets are represented as two independent rows with `a' being added to the bluer half and `b' being added to the redder half. The solid black line denotes the best fit to quiescence while the dashed line and colored data points represent the best flaring fit and flaring data respectively. The velocity offset from the laboratory wavelength is included in each plot as well.}
    \label{fig:flare_ion_fits}
\end{figure*}

For some of the brightest flares, select emission lines of interest are fit using the model emission line model described in section \ref{subsubsection:emission line model} to be a pseudovoigt profile with no linear component to represent the local continuum. Bright flares analyzed include 3, 5, 6, 11, 18, 20, and 21 all labeled in Figure \ref{fig:white-light_lc}. We analyzed the brightest ions in the data including Si IV, Si III, N V, C II, and Fe XXI. Figure \ref{fig:flare_ion_fits} depicts each of the chosen emission lines over each chosen flares in velocity space with best fits to both the quiescent data and specific flare data. Most flares include emission features being blue shifted from quiescence indicating the hot plasma moving toward the observer and away from the stellar surface. The blue shifted velocity offsets range from $-29\pm3$\,\kms{} to $-1.3\pm1$\,\kms{} with respect to quiescence. In addition, we find few emission lines red shifted from quiescence on the order of a few \kms{} potentially indicating emitting plasma falling back into the star. Particularly, 17 flares exhibit only blue shifts over each ion, 5 flares exhibit one red shifted line, 3 flares exhibit two red shifted lines, and 1 flare exhibits four red shifted lines. No flares show only red shifted lines. Typically, this red shifted line is Fe XXI which generally has a large uncertainty due to low signal. Si IV, N V, and Si III appear red shifted at least once while C II never appears red shifted.

\begin{figure}[h]
    \centering
    \includegraphics[width=0.9\linewidth]{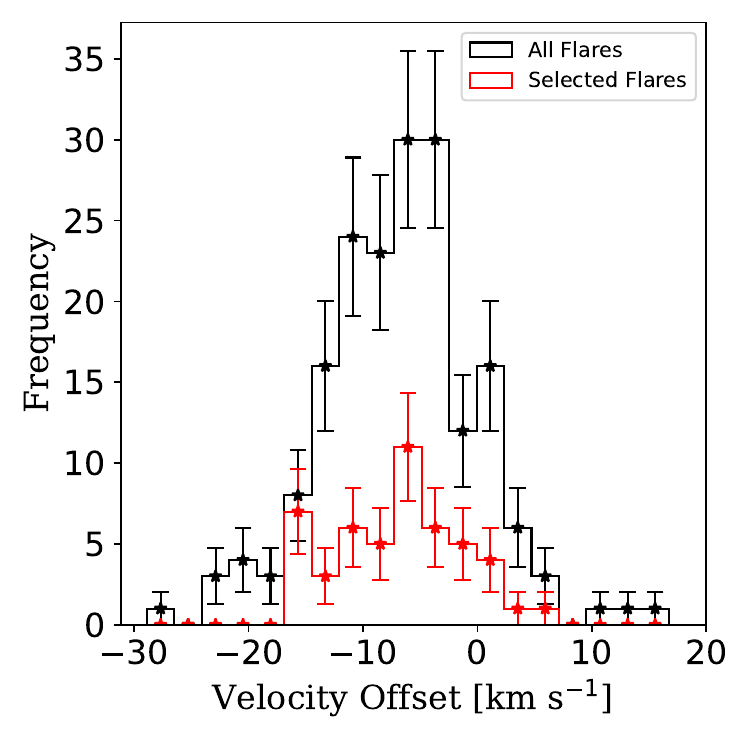}
    \caption{Histogram containing calculated velocity offsets for flaring data over all flares (in black) and selected flares (in red). Velocity offsets depict wavelength difference between flaring and quiescent spectra of their respective flares.}
    \label{fig:vel_hist}
\end{figure}

We take each of the calculated velocity offsets and plot them as a histogram to see if there exists a clear distribution of velocities. We calculate both velocity shifts for the selected flares as displayed in Figure \ref{fig:flare_ion_fits} as well as for all detected flares to see if the selection of flares itself has an impact on the distribution. The histogram can be found in Figure \ref{fig:vel_hist}. Based on this histogram, we see there is a clear normal distribution with a trend towards negative velocities. The distribution of only the selected flares has an average value of $-6.9 \pm 0.5$\,\kms{} while the distribution containing all flares has an average of $-6.9 \pm 0.8$\,\kms{}.

DS Tuc A has a large projected rotational velocity of $v\,sin\,i \approx27$\,\kms{} \citep{benatti_vsini}, such that spatially localized flare emission could acquire an apparent Doppler shift due to stellar rotation. The mean velocity offset measured ($\sim-7$\,\kms{}) corresponds to only $\approx{25\%}$ of rotational velocity indicating a modest longitudinal asymmetry could reproduce the small systematic bias seen if flares preferentially occurred on the approaching side of the star. However, with a rotation period of only $\approx8$\,days \citep{benatti} and less than half a day of COS observations, out dataset only samples $\approx{6\%}$ of a full rotation. Producing a persistent net blueshift under these conditions would require multiple independent flares across separate visits to originate from a stable, narrowly confined active longitude within the limited rotational phase observed. While this configuration cannot be entirely excluded, the combination of the limited rotational phase coverage and the presence of both redshifted and blueshifted measurements suggests rotation alone is unlikely to account for the observed velocity distribution.

The distribution of velocity offsets provides an insight into the kinematics of the upper stellar atmosphere during these flares. The prevalence of blueshifted lines across multiple ions and flares indicates upward moving plasma is a common outcome of the flaring process, even though the magnitude of these shifts varies greatly from lines to line \citep{blueshift_flares_doschek, blueshift_flares_feldman, blueshift_flares_antonucci}. We see no clear dependence on ion species or formation temperature as the velocities span similar ranges for each of the ions discussed (Si IV, Si III, C II, N V, Fe XXI) which suggests that flares are sampling a combination of upflows, downflows, and static material rather than a single coherent flow. The low, yet systematic offset towards negative velocities argues that on average, material heated by flaring is driven upwards. However, this shift being small compared to the overall spread of offsets implies contributions from stationary plasma as well as downflows may contribute to fitted profiles as well. The presence of a few red-shifted measurements points to episodes where cooling material is condensing back toward the stellar surface, or at least where localized downflows dominate the emission line. Notably, we do not detect bulk velocity motions of several hundred \kms{} as has been recently detected on the young solar analog EK Draconis and may be tracing a coronal mass ejection on that star~\citep{namekata_2026}.

The close agreement of average shift between the selected brightest flares and all flares indicates that the behavior as a whole is not strongly biased by flare brightness as both sets trace the same kinematic pattern. The resulting velocity distribution seen reflects intrinsic diversity of flare-induced motion in the upper stellar atmosphere with a preference towards upward flows. 

\subsection{A Flaring Continuum}

For identifying continuum regions in the spectrum, we wrote a short \texttt{python} script to simply read along the spectrum and identify segments of continuum based on the flux density surpassing a given threshold. The threshold is given as a multiple of the standard deviation of the master quiescent spectrum to find emission features in a matter similar to that of the flare finding algorithm described previously. After this, we do a cross check to ensure none of the tabulated emission lines in our line list (Appendix \ref{appendix:line_list}, Table \ref{tab:line_catalog}) appear within any of the continuum regions. This is done to ensure very small emission lines do not make it through the continuum thresholding. Identified and utilized continuum regions are written explicitly in Appendix \ref{appendix:continuum_regions}. We identify 47 segments of continuum totaling 116.7\,\AA{} of the total observed bandpass.

\begin{figure*}
    \centering
    \includegraphics[width=0.9\linewidth]{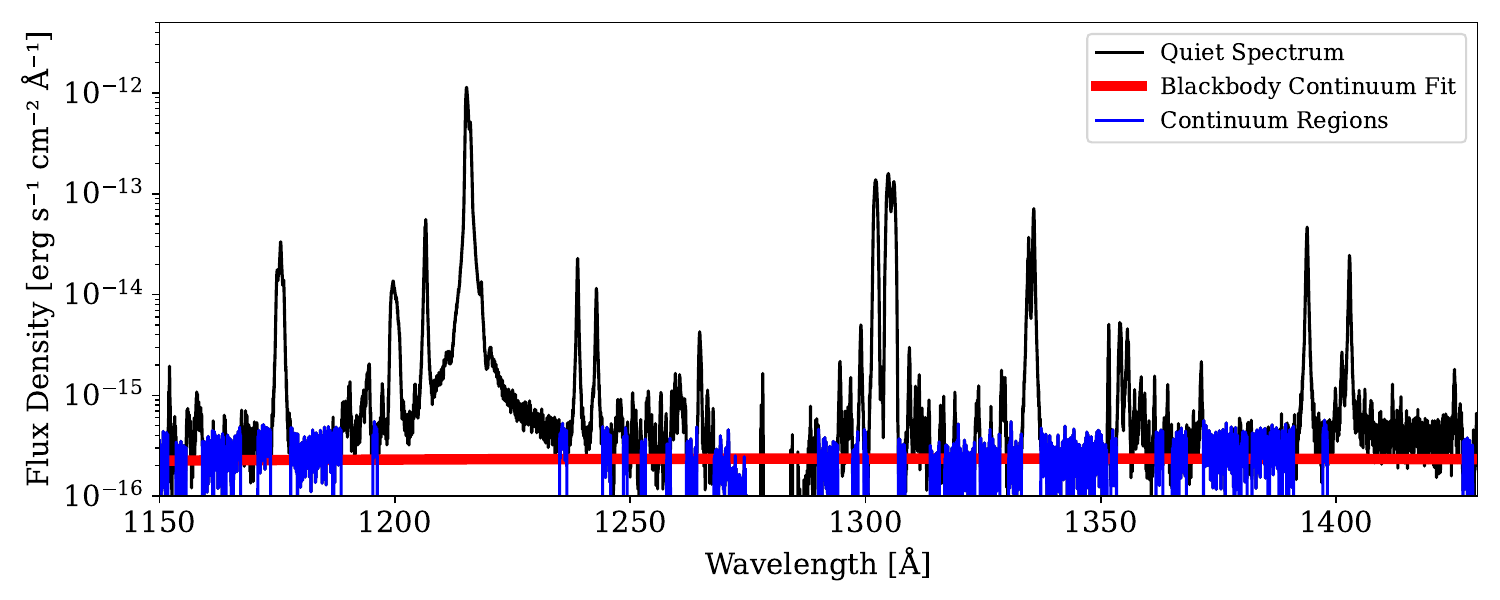}
    \caption{Blackbody best fit to quiescent spectrum continuum.}
    \label{fig:bb_fit}
\end{figure*}

We then fit blackbody models to the quiescent spectrum as well as the spectrum for each flare to calculate temperature and investigate the continuum response to flaring. The blackbody model was allowed to vary both the temperature and the scaling factor analogous to the size of the emitting region. The python module \texttt{scipy} is used for all continuum fitting and \texttt{scipy} provided covariance matrices are used for finding the uncertainty in the blackbody temperature. The average quiescent spectrum fit best with a blackbody temperature of $21650 \pm 279$\,K while the flaring spectra fit best with blackbody temperatures ranging between 15,000\,K to 32,000\,K however at such short durations, the errors reached 38\% in some flares. Taking maximum and minimum uncertainties into account, the best fit blackbody temperatures range about $12,000-40,000$ K. The best fit for the average quiescent spectrum is presented in Figure \ref{fig:bb_fit}. The average flaring spectra yielded a blackbody temperature of 22,000 $\pm$ 1000\,K. The best fit temperatures as well as emitting radius values estimated using the blackbody scaling coefficient are presented in Figure \ref{fig:bbfit_params}. See \ref{appendix:flares_bb_fits} for the best fit blackbody continuum fit for a select few flares covering varying energies and durations. As is clear by the large uncertainties, the fits are very poorly constrained for several flares including 5, 8, 9, and 14. In addition, none of the fits resulted in a reduced $\chi^2$ value of 1 as they were all far too small due to large uncertainties at the low continuum level as well as rather short flare durations. For these reasons, we draw no strong conclusions from these data.

In addition, none of the fits resulted in a reduced $\chi^2$ value near unity. Instead, the reduced $\chi^2$ values were substantially below unity, primarily because of the large uncertainties associated with the low continuum flux levels and the relatively short flare durations. Consequently, the $\chi^2$ statistic provides limited constraints on the quality of the blackbody fits, and we draw no strong conclusions regarding the physical interpretation of these fits.


\begin{figure}
    \centering
    \includegraphics[width=0.9\linewidth]{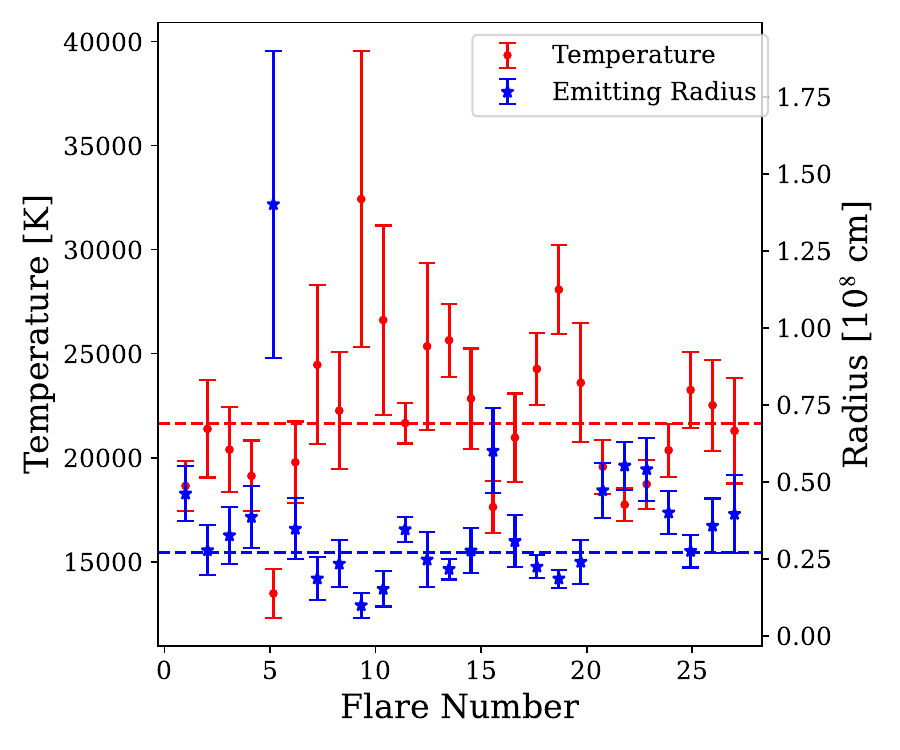}
    \caption{Best fit blackbody model parameters to flaring spectra. Left axis depicts blackbody temperature with data in red and right axis depicts emitting radius calculated using the scaling coefficient on the blackbody in blue. Dashed lines represent best fit values for quiescence.}
    \label{fig:bbfit_params}
\end{figure}

\subsubsection{Comparison with M Dwarfs}

\cite{froning_paper} analyzed a massive ($\delta > 30,000$ s) FUV flare on the M dwarf GJ 674 observed with \textit{HST} COS G130M covering the slightly bluer 1065\,\AA{} -- 1365\,\AA{} wavelength band. Similar to the work here, \cite{froning_paper} fit a single blackbody model to their FUV continuum to find a color temperature of T $\simeq 40,000 \pm 10,000$ K. While higher than the average blackbody temperature found for the DS Tuc A flares, this M dwarf color temperature is roughly consistent with several of the flares found on DS Tuc. On the contrary, work done by \cite{feinstein_paper} analyzes several flares from the M dwarf AU Microscopii with equivalent durations ranging $\delta {\approx}$ 1.8--688.5\,s with best fit blackbody temperatures of $T{\approx} $14,900--15,700\,K. While lower than the average blackbody temperature found for the DS Tuc A flares, we see nearly a third of the flares are within a few thousand kelvin of the AU Microscopii flares. This indicates the young solar analog DS Tuc A emits flares similarly in strength and variability as some M dwarfs. However, \citet{feinstein_paper} finds a sharp rise in flux blueward of 1100\,\AA{} that is not seen in the DS Tuc data analyzed here, likely due to the redder wavelength coverage in this work.

\subsection{Ion Contributions to Flares}

The contribution to each flare from each of the ions of interest is tabulated in Table \ref{tab:linesOfInterest}. These ions were selected as they emit several of the brightest lines in the 1130\,\AA{} to 1450\,\AA{} bandpass and each line serves to probe plasma of a specific temperature. These lines cover a decent distribution of FUV emitting plasma temperatures ranging from log(T) $\approx$ 4.4 with C II to log(T) $\approx$ 7.1 with Fe XXI. Specific temperatures for each ion can be found tabulated in Table \ref{tab:linesOfInterest}. The following sections have ions grouped into relative amounts of contribution to the total emitted energy, organized from highest to lowest contribution. Each section will discuss the ion's flaring to quiescent ratio and contribution to flares; a table of the average of these values of all flares can be found in Table \ref{tab:flare_quiet_ratios} for reference. Figure \ref{fig:flare18_lc} denotes flare 18 in each of the different ions for comparison.

\subsubsection{Si IV and C III Contribution}
Si IV and C III contribute most strongly to each of the detected flares (see also \citealt{namekata_2026}) with average contributions of $25.6 \pm 3.9$\% and $22.3 \pm 5.7$\% respectively. Both of these ions present clearly in each of the detected flares as they were the ions chosen for flare detection. The strong Si IV emission observed during flares indicates intense heating to transition-region temperatures, while the slightly weaker C III emission (and Si III emission we will briefly mention next) indicates that this heating extends into the warmer, upper layers of the transition region. This emission therefore traces plasma that has been heated to transition-region temperatures by energy deposited lower in the atmosphere, consistent with energy injection along magnetic field lines and subsequent collisional heating by energetic electrons as they precipitate into the chromosphere.

\subsubsection{Si III and C II Contribution}
Si III and C II contribute considerably to detected flares, however not nearly as strongly as their hotter counterparts Si IV and C III. On average, Si III contributes $15.2 \pm 5.3$\% of the observed FUV flux while C II contributes slightly less at $14.4 \pm 4.5$\%. Both Si III and C II have a noticeable presence in flares 1, 2, 5, 11, 14, 18, 19, 20, and 21 while Si III presents more strongly in flares 15, 16, and 24 and C II presents more strongly in flares 3, 7, 25, and 26 totaling 12 flares for Si III and 13 for C II. Unlike what is seen with Si IV, C III, and Si III, C II exhibits a comparatively low flare to quiescent ratio. This suggests the flare response in the cooler, lower transition region and upper chromospheric layers traced by C II is comparatively weaker, consistent with radiative hydrodynamic models in which energy deposition occurs mainly in the chromosphere and drives strong heating and evaporation into the upper transition region \citep[e.g.][]{allred_flaremodel1, allred_flaremodel2, reep}. In such models, plasma at C II formation temperatures may be rapidly heated to higher ionization states or contribute less prominently to the emergent flare spectrum compared to hotter transition region lines.



\subsubsection{N V and O V Contribution}
Compared to Si and C, N V is much weaker and is only noticeable in very strong flares while O V is practically flat with a hardly noticeable bump at the peak of the flare. We expect N V to be weaker as it forms at 200,000\,K, tracing less dense  upper chromosphere/lower transition region plasma than the formation temperature and density of Si IV and C III ions. O V similarly forms at a much hotter temperature $\sim{250,000}$\,K Significant flaring with N V can be found in flares 18, 20, and 21. A careful by-eye analysis shows there is are considerable bumps in the N V light curve corresponding to flares 1, 2, 5, 7, 11, and 18. O V is not seen significantly in any flare.

\subsubsection{A lack of Fe XXI Contribution}
\begin{figure}
    \centering
    \includegraphics[width=0.95\linewidth]{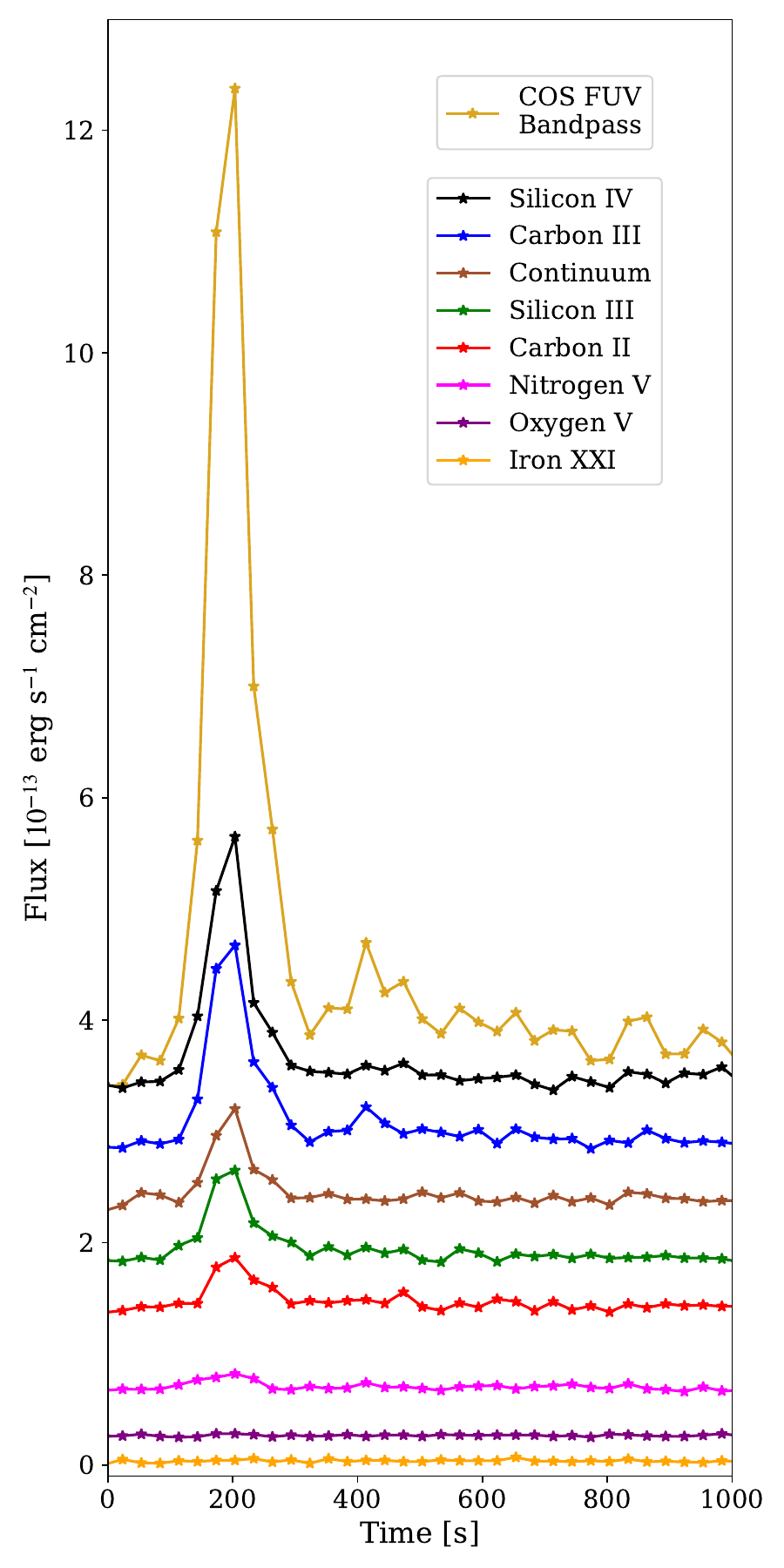}
    \caption{Light curves of flare 18 over each ion of interest. Individual light curves are positioned using a vertical offset from top to bottom in the order given by the legend. X-axis time depicts an arbitrary segment of time chosen to encompass the entire flare plus quiescence on both sides of the peak flux. All light curves are taken in 30 second time cadence except for Fe XXI and O V which were taken in 100 second time cadence due to relatively low signal. With the exception of the full COS FUV and Fe XXI light curves which have no vertical offset, the vertical offsets utilized are (from Si IV downward) 3, 2.5, 2, 1.5, 1, 0.5, 0.25 in units of $10^{-13}$ erg s$^{-1}$ cm$^{-2}$.}
    \label{fig:flare18_lc}
\end{figure}

As can clearly be seen in the bottom two plotted lines in Figure \ref{fig:flare18_lc}, even in the most energetic flare Fe XXI (probing plasma with temperatures around $12,000,000$\,K) remains completely flat. We did not detect any enhancements in this line at the flare peak above the continuum.
To explain this, we can first look at the scaling relation presented as equation 2 in \citet{ayres_fexxi_bol} comparing bolometric luminosity, X-ray luminosity, bolometric flux, and Fe XXI flux. From \cite{newton} we know DS Tuc A has a bolometric flux of $(1.2026\pm0.017)\times10^{-8}$\,erg\,s$^{-1}$\,cm$^{-2}$ and from \cite{chandra_flares} we know DS Tuc A has an $L_{\text{X-ray}}/L_{bol}$ of $4.74\times10^{-4}$. Using these values we can calculate the DS Tuc A Fe XXI flux to be $\approx{2.82\times10^{-15}}$\,erg\,s$^{-1}$\,cm$^{-2}$. This is only 60\% of the measured Fe XXI quiescent flux and 70\% of the measured Fe XXI flux in flare 18. Using typical flare emission measurements and CHIANTI emissivity for Fe XXI (1354\AA{}), wse can assess the range of peak line intensity for flare class X1 as $10^{-2}$\,erg\,s$^{-1}$\,cm$^{-2}$, class X10 as $10^{-1}$\,erg\,s$^{-1}$\,cm$^{-2}$, and class X100 as 1\,erg\,s$^{-1}$\,cm$^{-2}$. At DS Tuc A's distance of $9.1\times10^6$\,AU, we can translate these to $1.2\times10^{-16}$\,erg\,s$^{-1}$\,cm$^{-2}$ for class X1, class X10 as $1.2\times10^{-15}$\,erg\,s$^{-1}$\,cm$^{-2}$, and class X100 as $1.2\times10^{-14}$\,erg\,s$^{-1}$\,cm$^{-2}$. To uncover noticeable Fe XXI flaring, we would need to see hotter, more energetic flares.


\subsubsection{Continuum Contribution}
The continuum on average contributes similarly to flares as Si III with a contribution of $14.2 \pm 3.8$\%. However, the continuum responds much more clearly during flaring with small but clear presence in each of the detected flares. 

\begin{table}[h]
    \centering
    \setlength{\tabcolsep}{3pt}
    \begin{tabular}{c|cc}
       Ion  & Average Ratio (F/Q) & Average Contribution \\
       \hline \hline
       C II   & 1.091 $\pm$ 0.032 & 14.4 $\pm$ 4.5 \% \\
       Si III & 1.117 $\pm$ 0.050 & 15.2 $\pm$ 5.3 \% \\
       C III  & 1.263 $\pm$ 0.052 & 22.3 $\pm$ 5.7 \% \\
       Si IV  & 1.397 $\pm$ 0.033 & 25.6 $\pm$ 3.9 \% \\
       N V    & 1.132 $\pm$ 0.033 & 2.5  $\pm$ 1.2 \% \\
       O V    & 1.155 $\pm$ 0.078 & 1.6  $\pm$ 0.44 \% \\
       Fe XXI & 1.057 $\pm$ 0.067 & 1.1  $\pm$ 0.25 \% \\
       Continuum & 1.209 $\pm$ 0.052 & 14.2 $\pm$ 3.8 \%
    \end{tabular}
    \caption{Average flaring to quiescent line ratios and average ion contribution to each flare.}
    \label{tab:flare_quiet_ratios}
\end{table}

\subsection{Electron Density of the Flaring Regions}
The Si IV ($\lambda1393.76$) / O IV ($\lambda1401.16$) serves as a diagnostic of electron density in the stellar transition region \citep{siIV_oIV_ratios}. By finding the ratio of these emission lines over the duration of the flare, we can estimate an average electron density over the course of the flare. For this, only flares with a duration longer than 200 seconds were utilized as at durations less than this, the signal in the O IV line is too low. We subtract the quiescent flux from each flare flux before calculating the line ratios to ensure we only include flaring emission. Work done by \cite{elec_dens_paper} models the electron density at a temperature of log(T/K) = 4.88 as a function of the Si IV / O IV line ratio. We interpolate this model to find the electron densities corresponding to our measured line ratios. We interpolate using the \texttt{SciPy PchipInterpolator} which fits continuous, smooth, piecewise cubics from point to point in the model. Calculated line ratios and electron densities are presented in Table \ref{tab:elec_densities}. Flare 15 is the only flare with a duration greater than 200 seconds that was not included as it's line ratio fell below the provided model thus did not fit within the interpolation; no extrapolation is done for this work.

\begin{table}[h]
    \centering
    \begin{tabular}{c|cc}
        Flare & Si IV / O IV & Electron Density \\
         & Ratio & [log(N$_e$ / cm$^{-3}$)] \\
        \hline \hline
        1 & 5.77 $\pm$ 0.85 & 10.98 $^{+0.14} _{-0.21}$ \\
        2 & 4.26 $\pm$ 1.40 & 10.46 $^{+0.50} _{-1.46}$ \\
        3 & 6.08 $\pm$ 1.82 & 11.03 $^{+0.25} _{-0.58}$ \\
        4 & 4.67 $\pm$ 1.74 & 10.67 $^{+0.42} _{-1.67}$ \\
        6 & 5.89 $\pm$ 2.34 & 11.00 $^{+0.31} _{-2.00}$ \\
        7 & 1.57 $\pm$ 1.04 & 9.00 $^{+0.00} _{-0.00}$ \\
        8 & 1.85 $\pm$ 2.07 & 9.00 $^{+1.04} _{-0.00}$ \\
        9 & 2.22 $\pm$ 2.45 & 9.00 $^{+1.67} _{-0.00}$ \\
        10 & 1.61 $\pm$ 2.42 & 9.00 $^{+1.22} _{-0.00}$ \\
        11 & 11.53 $\pm$ 0.80 & 11.55 $^{+0.04} _{-0.04}$ \\
        13 & 5.55 $\pm$ 0.71 & 10.94 $^{+0.13} _{-0.20}$ \\
        14 & 4.02 $\pm$ 1.05 & 10.22 $^{+0.60} _{-1.22}$ \\
        16 & 5.68 $\pm$ 0.96 & 10.96 $^{+0.16} _{-0.27}$ \\
        17 & 11.28 $\pm$ 1.13 & 11.53 $^{+0.06} _{-0.06}$ \\
        18 & 30.75 $\pm$ 1.23 & 12.09 $^{+0.02} _{-0.02}$ \\
        19 & 4.57 $\pm$ 1.11 & 10.62 $^{+0.34} _{-1.62}$ \\
        20 & 8.31 $\pm$ 0.82 & 11.32 $^{+0.07} _{-0.09}$ \\
        21 & 36.76 $\pm$ 1.96 & 12.18 $^{+0.03} _{-0.03}$ \\
        22 & 10.92 $\pm$ 1.47 & 11.51 $^{+0.08} _{-0.09}$ \\
        23 & 5.04 $\pm$ 0.99 & 10.81 $^{+0.22} _{-0.53}$ \\
        24 & 7.69 $\pm$ 1.05 & 11.26 $^{+0.11} _{-0.14}$ \\
    \end{tabular}
    \caption{Measured ratios of the Si IV ($\lambda$1393.76) to O IV ($\lambda$1401.16) emission lines with corresponding electron density measurements.}
    \label{tab:elec_densities}
\end{table}

The derived Si IV/O IV ratios span nearly two orders of magnitude, resulting in electron densities from log($N_e$/cm$^{-3}$) $\approx{}$ 10.2 to 12.1. This broad range indicates that the transition-region plasma conditions vary substantially across flares, with several events reaching densities characteristic of strongly compressed or rapidly heated material. Most flares cluster near log($N_e$) $\approx{}$ 11, suggesting that densities around $10^{11}$ cm$^{-3}$ represent typical conditions in the emitting regions during moderate events. This is consistent with densities found in EK Draconis using the same diagnostic lines \citep{ayres_fexxi_bol}. The highest ratios---seen in flares 18, 21, and to a lesser degree 10 and 17---correspond to densities above log $N_e$ $\approx{}$ 11.5, implying episodes of particularly intense energy deposition or efficient confinement that produces unusually dense transition-region plasma. \citet{solar_TR_edens} use the OIV] 1401.16\,\AA{} line as a density diagnostic for measuring electron density of transition region loops in the Sun. From their work, they derive electron densities on the order of $10^{10}-10^{11}\,\text{cm}^{-3}$ which agrees well with the electron densities calculated for the DS Tuc flares. 

The larger uncertainties on the lower-ratio flares reflect the reduced sensitivity of the diagnostic at smaller Si IV/O IV values, yet even these cases consistently indicate densities well above solar quiescent transition-region levels (typically $\approx{10^{10}}$ cm$^{-3}$ \citep{sun_edens}), reinforcing the idea that flare heating universally drives substantial compression. This distribution of densities reveals a population of flares with broadly similar plasma conditions with a subset of exceptionally dense, high-ratio events, emphasizing the diversity of environments produced during flaring on this young solar analog.

\subsection{Bolometric Energy of DS Tuc A Flares}
\begin{figure}
    \centering
    \includegraphics[width=0.99\linewidth]{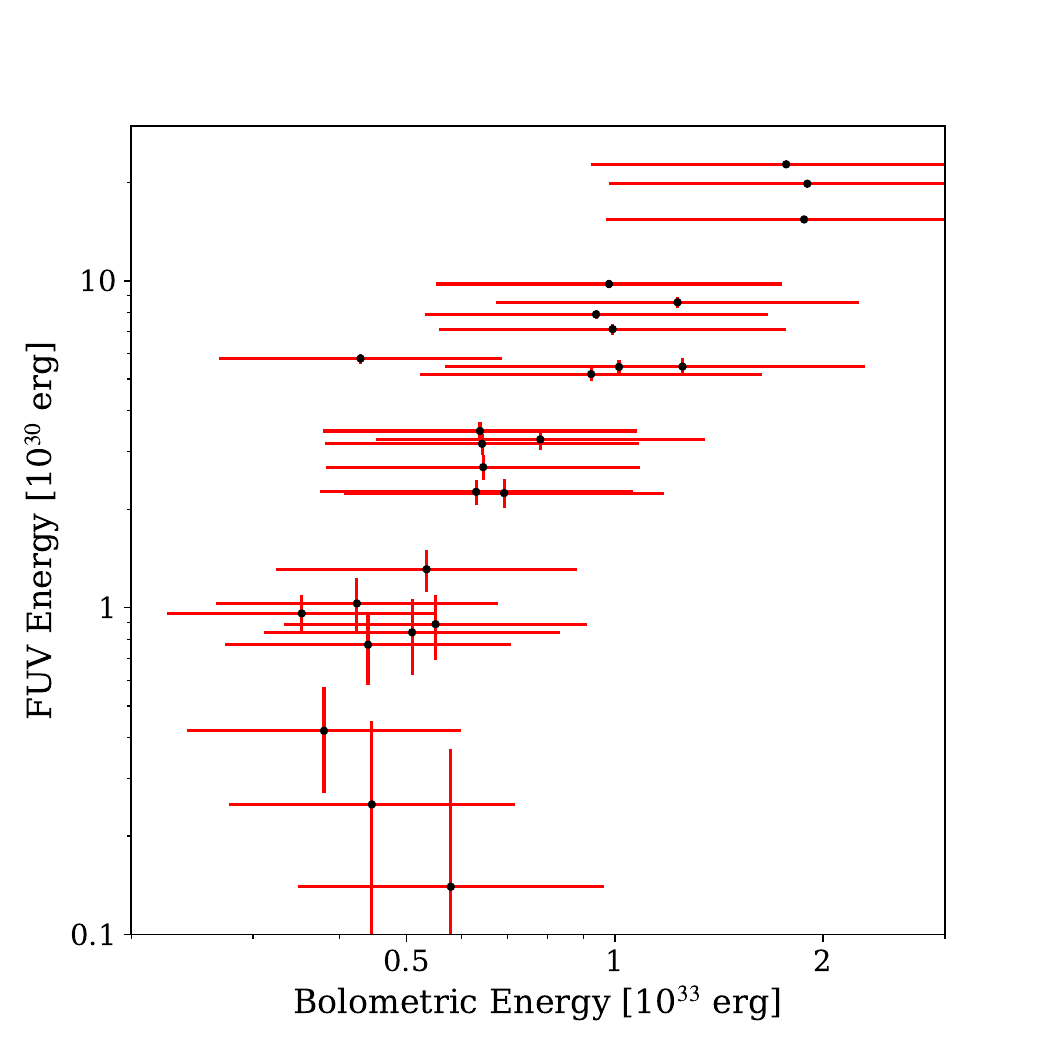}
    \caption{FUV energy versus scaled bolometric energy.}
    \label{fig:FUV_vs_bolometric}
\end{figure}

There exists an empirical relationship between the 977\,\AA{} C III and emitted X-ray flux from a flare \citep{namekata_2026}. 
As this line is not within our \textit{HST} data bandpass, we use the C III multiplet at 1176\,\AA{} and line ratios found between the two C III line in \cite{CIII_ratio} for various young stars. While the work does not find a ratio for DS Tuc, we take an average of the ratios for other similar stars including the M0 star AU Microscopii, the K1V star AB Doradus, and the K2V star $\epsilon$ Eridani. Taking an average of the ratios, we adopt the C III 1176\,\AA{}/977\,\AA{} ratio for DS Tuc A as 0.74 $\pm$ 0.01. 
Likewise, \cite{namekata_2026} scaled the C III 1176 {\AA} to C III 977 flux by using the flux ratio of 0.66 which is obtained by CHIANTI model, and our new estimate is very consistent with previous approach.
Using the ratio of 0.74, we convert C III 1176\,\AA{} flux to C III 977\,\AA{} flux and then use the GOES X-ray (1\,\AA{}-8\,\AA{}) flux scaling relation to convert C III 977\,\AA{} energy into GOES X-ray peak flux. 
From work on solar flares completed by \cite{xray_to_bolo1} and \cite{xray_to_bolo2}, we know X-ray (1\,\AA{}-8\,\AA{}) is, on average, approximately 1\% of the flare bolometric energy. \cite{goes_to_bolo} provide an empirical relationship between GOES peak X-ray flux and bolometric energy which we use to calculate bolometric energies for the DS Tuc A flares. Peak GOES X-ray fluxes and scaled bolometric energies for each flare are presented in Table \ref{tab:scaled_energies} along with the FUV energy calculated in \ref{sec:ffds} for comparison. Here, the inferred bolometric flare energies range from 0.35--1.9 $\times10^{33}$\,erg, with six events having inferred energies above the $10^{33}$\,erg threshold. A superflare is conventionally defined as a flare releasing at least $10^{33}$\,erg in bolometric energy \citep{superflare_def}. Flares 11, 13, 17, 18, 21, and 23 therefore qualify as superflare candidates. We refer to these events as candidates rather than confirmed superflares because their bolometric energies are inferred using empirical solar-flare scaling relations and are consequently subject to substantial systematic uncertainties.
A plot of FUV absolute energy over the observed bandpass versus scaled bolometric energy is presented in Figure \ref{fig:FUV_vs_bolometric}. 


\begin{figure}
    \centering
    \includegraphics[width=0.99\linewidth]{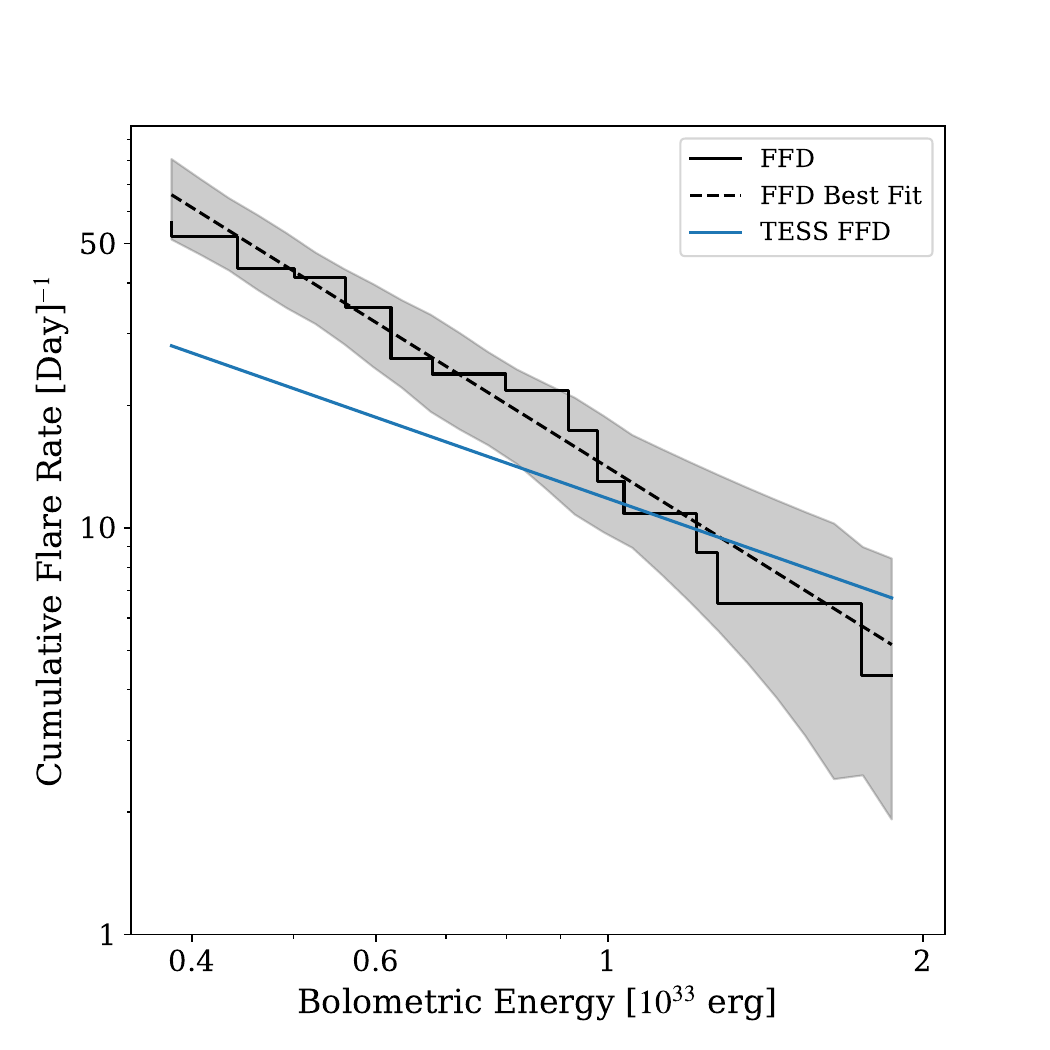}
    \caption{Bolometric Energy FFD with \textit{TESS} FFD overlaid.}
    \label{fig:ffd_w_tess}
\end{figure}

We construct an FFD identically following the process of section \ref{sec:ffds} using the calculated bolometric energies. We fit this FFD with a power law and find an index of $1.5\pm0.1$. The FFD is presented in Figure \ref{fig:ffd_w_tess} with the \textit{TESS} FFD from \citet{TESS_FFD} overlaid, with the same normalization. Figure 6 in \cite{TESS_FFD} presents the flare frequency distributions for the three \textit{TESS} sectors of DS Tuc A data available. It is evident that flare cumulative frequency between sectors varies by a factor of 2 to 3. This can be explained by the presence of flare-productive active regions \citep{toriumi_active_flare_region}. that boost the flare frequency over the lifetime of a large active region that can last for weeks. Also, \textit{TESS} data provide an upper limit for superflare frequency from DS Tuc A as the \textit{TESS} spatial resolution cannot resolve flares from DS Tuc B, which is also an active K type star. If we account for such variability, our results appear to be consistent with variable frequency of flares observed in the white-light band.

\begin{table}[]
    \centering
    \setlength{\tabcolsep}{2pt}
    \begin{tabular}{l|lll}
        Flare & FUV Energy & GOES X-ray Peak & Bolometric \\
         & [$10^{30}$ erg] & Flux [erg\,s$^{-1}$\,cm$^{-2}$] & Energy [$10^{33}$ erg] \\
        \hline \hline
        1 & 9.77 $\pm$ 0.29 & 12.68 $\pm$ 1.38 & 0.98 $^{+0.76}_{-0.43}$ \\
        2 & 3.47 $\pm$ 0.23 & 7.37 $\pm$ 1.07 & 0.64 $^{+0.44}_{-0.26}$ \\
        3 & 3.17 $\pm$ 0.23 & 7.43 $\pm$ 1.10 & 0.64 $^{+0.44}_{-0.26}$ \\
        4 & 2.69 $\pm$ 0.24 & 7.46 $\pm$ 1.02 & 0.65 $^{+0.44}_{-0.26}$ \\
        5 & 0.77 $\pm$ 0.19 & 4.60 $\pm$ 0.93 & 0.44 $^{+0.27}_{-0.17}$ \\
        6 & 0.14 $\pm$ 0.23 & 6.51 $\pm$ 0.96 & 0.58 $^{+0.39}_{-0.23}$ \\
        7 & 0.25 $\pm$ 0.20 & 4.67 $\pm$ 0.83 & 0.45 $^{+0.27}_{-0.17}$ \\
        8 & 0.84 $\pm$ 0.22 & 5.53 $\pm$ 0.89 & 0.51 $^{+0.32}_{-0.20}$ \\
        9 & 1.03 $\pm$ 0.20 & 4.38 $\pm$ 0.79 & 0.42 $^{+0.25}_{-0.16}$ \\
        10 & 5.78 $\pm$ 0.20 & 4.45 $\pm$ 0.82 & 0.43 $^{+0.26}_{-0.16}$ \\
        \hline
        11 & 19.80 $\pm$ 0.41 & 29.24 $\pm$ 2.62 & 1.90 $^{+1.77}_{-0.92}$ \\
        12 & 0.96 $\pm$ 0.13 & 3.48 $\pm$ 1.24 & 0.35 $^{+0.20}_{-0.13}$ \\
        13 & 8.59 $\pm$ 0.33 & 16.93 $\pm$ 1.71 & 1.23 $^{+1.02}_{-0.56}$ \\
        14 & 2.26 $\pm$ 0.20 & 7.25 $\pm$ 1.24 & 0.63 $^{+0.43}_{-0.26}$ \\
        15 & 2.24 $\pm$ 0.23 & 8.15 $\pm$ 1.18 & 0.69 $^{+0.49}_{-0.29}$ \\
        16 & 3.27 $\pm$ 0.24 & 9.50 $\pm$ 1.29 & 0.78 $^{+0.57}_{-0.33}$ \\
        17 & 5.46 $\pm$ 0.34 & 17.29 $\pm$ 6.29 & 1.25 $^{+1.05}_{-0.57}$ \\
        18 & 22.70 $\pm$ 0.34 & 26.75 $\pm$ 2.75 & 1.77 $^{+1.62}_{-0.85}$ \\
        19 & 0.89 $\pm$ 0.20 & 6.11 $\pm$ 1.10 & 0.55 $^{+0.36}_{-0.22}$ \\
        20 & 7.89 $\pm$ 0.24 & 12.01 $\pm$ 1.65 & 0.94 $^{+0.72}_{-0.41}$ \\
        \hline
        21 & 15.40 $\pm$ 0.42 & 28.85 $\pm$ 2.59 & 1.88 $^{+1.75}_{-0.91}$ \\
        22 & 5.18 $\pm$ 0.25 & 11.76 $\pm$ 1.49 & 0.92 $^{+0.71}_{-0.40}$ \\
        23 & 5.45 $\pm$ 0.29 & 13.23 $\pm$ 1.48 & 1.01 $^{+0.80}_{-0.45}$ \\
        24 & 7.11 $\pm$ 0.28 & 12.88 $\pm$ 1.52 & 0.99 $^{+0.78}_{-0.44}$ \\
        25 & 1.31 $\pm$ 0.19 & 5.88 $\pm$ 1.14 & 0.53 $^{+0.35}_{-0.21}$ \\
        26 & 0.42 $\pm$ 0.15 & 3.82 $\pm$ 1.09 & 0.38 $^{+0.22}_{-0.14}$ \\
    \end{tabular}
    \caption{Bolometric flare energies. FUV flare energies are taken over the observed \textit{HST} COS bandpass, GOES Peak X-ray flux is from the CIII 977\,\AA{} relation, and bolometric flare energies are scaled from the GOES X-ray (1\,\AA{}-8\,\AA{}) flux.}
    \label{tab:scaled_energies}
\end{table}

The scaled energies reveal that the FUV bandpass only captures a small fraction of the total radiative output of each flare. While the FUV measured energies span about two orders of magnitude, the corresponding bolometric energies are consistently larger by factors of $10^2$-$10^3$, reflecting flare emission weighted towards hotter plasma and broad-band continuum emission. 
The inferred GOES-band X-ray peak fluxes, estimated from the C III 977\,\AA{} scaling relation, provide a complementary measure of the instantaneous flare intensity in the soft X-ray regime rather than a total radiated energy. These peak fluxes span roughly an order of magnitude across the flare sample and broadly track the relative flare strengths inferred from the FUV energies. 
The subset of the most energetic events--such as flares 10, 18, 21, and 23--show particularly elevated C III-based X-ray outputs, resulting in bolometric energies approaching or exceeding $10^{33}$ erg, consistent with large flares observed on young, rapidly rotating solar analogs \citep{solaranalog_xray}. 

The power-law index derived from the bolometric FFD constructed from the COS observations (1.5 $\pm$ 0.1) is somewhat steeper than that inferred from the \textit{TESS} bolometric FFD (0.9 $\pm$ 0.17) \citep{TESS_FFD}. While these values are not fully consistent, the difference is modest and may reflect systematic effects associated with the different methods used to infer bolometric flare energies from FUV and optical observations, as well as differences in sensitivity and flare detectability between the two datasets. This mild tension suggests that the FFDs derived from \textit{TESS} and FFDs derived from COS may sample slightly different portions of the underlying flare population, but does not preclude a similar statistical description of flare energies across the two approaches.

\section{Conclusion}
\label{sec:conclusion}

Our broad analysis of \textit{HST}/COS observations of DS Tuc A reveal a coherent picture of an active young solar analog whose flaring behavior is both energetically significant and dynamically complex. Across major transition-region ions, the FUV flare frequency distributions exhibit statistically similar power-law indices, indicating that the observed variability arises from a single underlying population of events rather than ion-specific flare classes. These indices are also consistent with those seen in M-dwarf FUV studies but appear flatter than DS Tuc’s own \textit{TESS} white-light FFD, implying a relatively lower occurrence of high-energy flares in the FUV compared to other bandpasses. 

The emission line velocity shifts measured during flares further emphasize the variability of dynamics in the system. 
Although DS Tuc A's large projected rotational velocity ($v\,sin\,i\approx{27}$\,\kms{}) could in principle introduce apparent Doppler shifts from localized emission, the modest magnitude of the mean offset and the limited rotational phase coverage suggest rotation alone is unlikely to account for the observed biased distribution. Blueshifted line centers nevertheless dominate most ions during flares, pointing to flare-driven upward motion in the transition region. The lack of a systematic trend with ion species, combined with a wide span of velocities and occasional redshifts, suggests that each flare samples a mixture of upflows, downflows, and static material rather than a single coherent flow. The consistency of this behavior across flare strengths reinforces the interpretation that the velocity shifts primarily reflect intrinsic atmospheric variability.

The relative flare contributions of individual ions highlight the response of the stellar atmosphere to magnetic energy release. Strong Si IV emission and comparatively weaker C III and Si III enhancements indicate that the bulk of the flare energy is deposited in the hotter, upper transition region, whereas the modest response of C II shows that significantly less energy reaches the cooler, lower layers. This pattern reflects the rapid heating and magnetic reconnection expected to occur high in the atmosphere of a young, magnetically active star.

Electron densities derived from Si IV/O IV ratios show that flare-driven plasma compression is substantial across the collected 26 flare sample. While most flares have densities of log$(N_e) \approx{11}$, several flares achieve densities greater than log$(N_e) \approx{11.5}$, demonstrating that some events generate exceptionally dense transition region conditions. Even the lower-density cases exceed typical quiet-Sun values, indicating that flare heating universally elevates plasma densities well above quiescent levels in this system.

Finally, scaling the FUV energies to bolometric values confirms that the COS bandpass captures only a small fraction of the total radiated energy. Bolometric estimates exceed the measured FUV energies by factors of $10^2$-$10^3$, with the largest flares exceeding $10^{33}$ erg, consistent with the energetic output expected from young, rapidly rotating solar analogs. The bolometric FFD slope inferred from the COS data is somewhat steeper than that derived from \textit{TESS} photometry, indicating a mild inconsistency between the two bolometric flare energy distributions. While the difference is modest, it suggests that bolometric FFDs constructed from FUV and optical observations may be subject to systematic effects associated with the adopted scaling relations and/or differences in flare detectability across bandpasses. As such, the COS-detected flares broadly trace the high-energy flare population of DS Tuc A, but direct comparison with the \textit{TESS}-derived bolometric FFD should be treated with caution.

In the future, we plan to perform coordinated observations with \textit{TESS}, X-ray, and \textit{HST} observations to derive the flare energy budget and the frequency of occurrence for DS Tuc A and B.


Ultimately, these results portray DS Tuc A as a highly active young solar analog whose flares heat, compress, and dynamically restructure its chromosphere and transition region while contributing significantly to its overall radiative output. The consistency of flare statistics across ions and wavelengths highlights the robustness of the observed flare population and provides a foundation for analyzing the impact of such activity on the star’s circumstellar environments, atmospheric escape, and magnetospheric activity of its exoplanet DS Tuc Ab.
Our results also provide a unique preview of the flare activity of the young Sun at the turbulent time on early Earth after the aftermath of the Moon-forming impact — still largely molten and enveloped in a rock-vapor atmosphere \citep{kumari2025}.


\begin{acknowledgments}

T.S., K.F., and V.S.A. acknowledge the financial support from DDT HST HST-GO-17305.002-A. V.S.A. was also supported by the NASA Planetary Science Division’s Internal Scientist Funding Model (ISFM), V.S.A. acknowledges support from the NASA/GSFC Sellers Exoplanet Environments Collaboration (SEEC), which is funded by the NASA Planetary Science Division’s Internal Scientist Funding Model (ISFM), NASA’s Astrophysics Theory Program grant \#80NSSC24K0776 and NASA’s Exoplanetary Research Program grant \#NNH21ZDA001N-XRP.

The HST data presented in this article were obtained from the Mikulski Archive for Space Telescopes (MAST) at the Space Telescope Science Institute. The specific observations analyzed can be accessed via \dataset[doi: 10.17909/gq4d-0024]{https://doi.org/10.17909/gq4d-0024}.

\end{acknowledgments}


\facilities{\textit{Hubble Space Telescope}}

\software{numpy \citep{numpy}, scipy \citep{scipy}, matplotlib \citep{matplotlib}, astropy\cite{astropy:2013, astropy:2018, astropy:2022}, \texttt{lmfit} \citep{lmfit}, \texttt{calcos}\footnote{https://github.com/spacetelescope/calcos}, \texttt{costools}\footnote{https://github.com/spacetelescope/costools}, \texttt{calcos/costools} \citep{cos_handbook}. 
}

\clearpage

\appendix

\section{Continuum Regions}
\label{appendix:continuum_regions}
\noindent
\small
[1124.209, 1133.139], [1136.989, 1138.319], [1143.319, 1145.349], [1145.989, 1151.829], [1153.519, 1155.688], [1159.158, 1161.338], [1161.548, 1163.688], [1164.118, 1167.278], [1170.888, 1173.648], [1177.898, 1186.818], [1187.028, 1188.628], [1195.347, 1196.287], [1235.016, 1236.576], [1244.136, 1245.886], [1248.546, 1249.376], [1252.356, 1253.226], [1257.766, 1258.566], [1261.966, 1264.176], [1290.000, 1294.106], [1297.256, 1298.406], [1299.656, 1300.586], [1306.965, 1308.405], [1313.695, 1315.695], [1316.815, 1318.775], [1324.325, 1326.495], [1326.835, 1328.555], [1330.375, 1333.125], [1337.225, 1351.304], [1352.104, 1353.324], [1361.684, 1363.164], [1365.114, 1366.364], [1366.574, 1368.154], [1371.774, 1378.293], [1378.503, 1379.313], [1379.753, 1381.063], [1382.083, 1385.473], [1385.823, 1387.853], [1388.493, 1389.463], [1389.883, 1390.953], [1397.133, 1398.113], [1426.882, 1429.022], [1430.382, 1437.242], [1437.842, 1440.572]

\section{Master Quiescent Spectrum}
\begin{figure*}[h]
    
\includegraphics[width=0.99\linewidth]{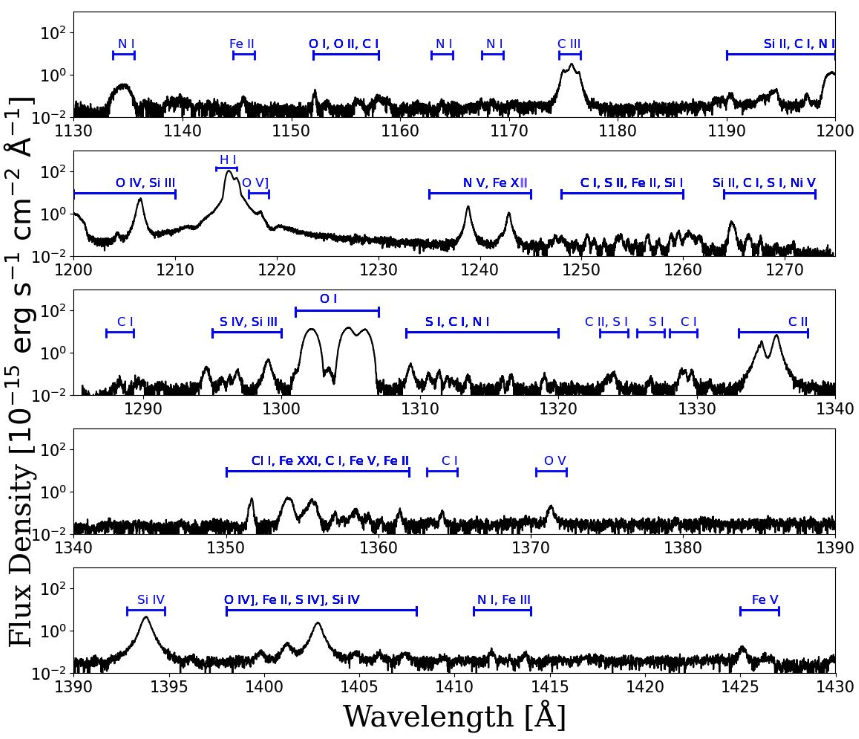}
    \caption{Master quiescent spectrum from DS Tuc \textit{HST}-COS data.}
    \label{fig:masterQuiet}
\end{figure*}

\clearpage

\section{Select Flaring Spectra Blackbody Fits}
\label{appendix:flares_bb_fits}
\begin{figure}[h]
    \centering
    \includegraphics[width=0.92\linewidth]{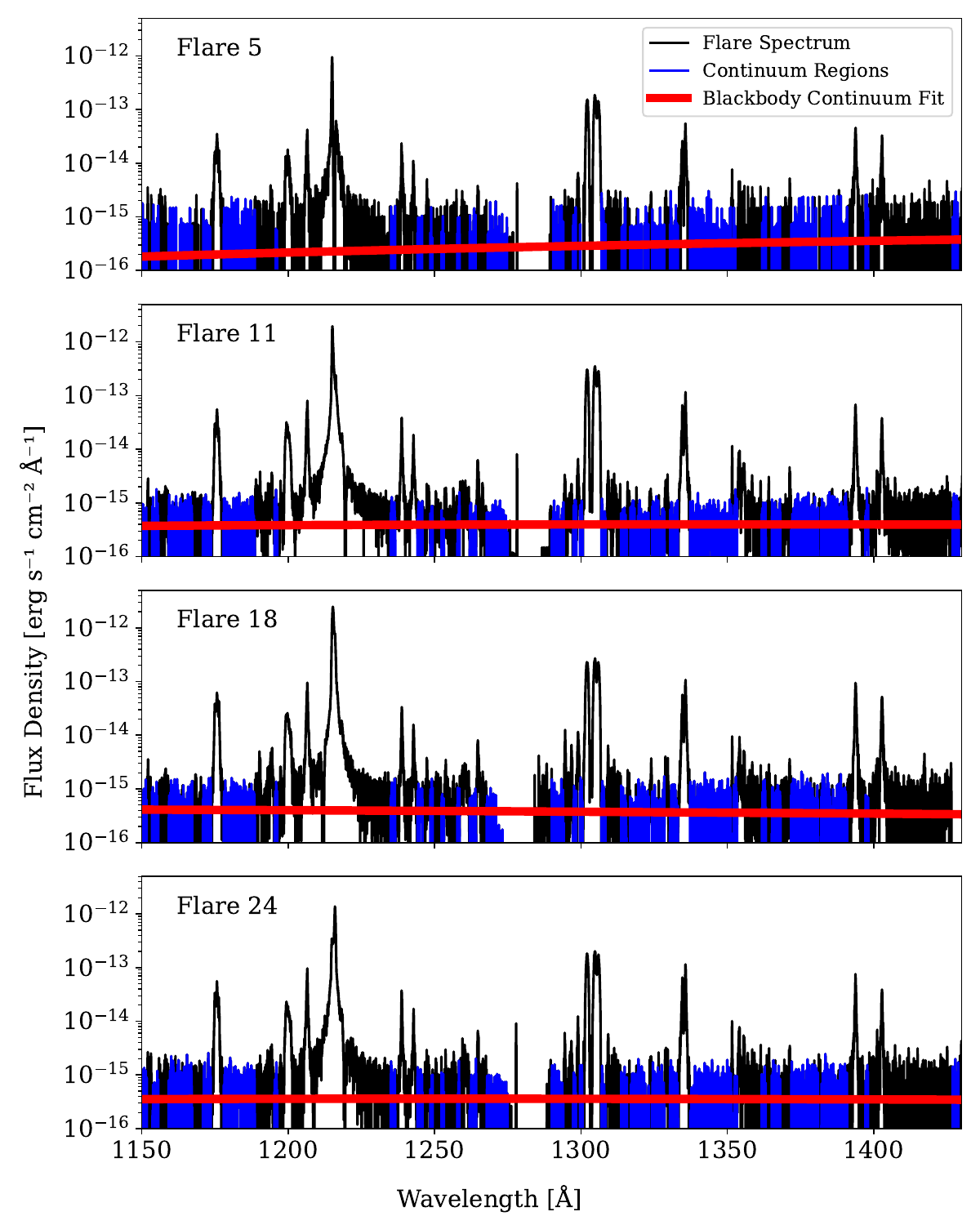}
    \caption{Best fit blackbody continuum to flares 5, 10, 18, and 24. Flaring spectrum denoted in black with the continuum regions overlaid in blue. The best fit is presented in red.}
    \label{fig:flares_bb_fits}
\end{figure}

\section{Line List}
\label{appendix:line_list}
\begin{table}[h!]
    \centering
\small
\setlength{\tabcolsep}{1.5pt}
\begin{tabular}{l|l|l|l|l|l|l|l|c}

       Ion    & $\lambda_{\text{rest}} [\textnormal{\AA{}}]$  & $\lambda_{\text{obs}} [\textnormal{\AA{}}]$ & V$_{\text{Shift}} [\text{km} \, \text{s}^{-1}]$ & Flux (Q) & Flux (F) & FWHM (Q) &   FWHM (F) & Notes\\

        \hline \hline
       
        N I  & $1134.415$ & $1134.586 \pm 0.038$ & $45.230 \pm 0.223$ & {$7.545 \pm 0.733 $}  & {$-$} & {$2.218 \pm 0.234$} &   {$1.132 \pm 0.159$}\\

        Fe II  & $1145.641 \pm 0.002$ & $1145.615 \pm 0.012$ & $-6.932 \pm 0.460$ & {$0.594\pm 0.028$}  & {$0.864 \pm 0.367$} & {$0.733 \pm 0.043$} &   {$1.025 \pm 0.566$}\\

       O I & $1152.151$ & {$1152.157 \pm 0.006$} & $1.407 \pm 1.123$ &  {$0.659 \pm 0.028$}&  {$1.031 \pm 0.122$} &{$0.399 \pm 0.021$} &   {$0.293 \pm 0.044$}\\

       O II  & $1153.368$ & $1153.224 \pm 0.025$ & $-37.518 \pm 0.175$ & {$0.598\pm 0.048$}  & {$-$} & {$1.180 \pm 0.117$} &   {$-$}& [1]\\

       C I & {$1156.186 \pm 0.015$} & {$1156.117 \pm 0.035$} & {$-17.916 \pm 0.553$} &  {$1.116 \pm 0.127$}&  {$0.907 \pm 0.222$} &{$1.740\pm 0.231$} &   {$0.951 \pm 0.312$}\\

      C I  & $1158.03 \pm 0.008$ & $1158.047 \pm 0.020$ & $4.350 \pm 1.264$ & {$1.152\pm 0.075$}  & {$1.174 \pm 0.183$} & {$1.213 \pm 0.098$} &   {$0.643 \pm 0.130$}\\

      N I  & $1163.884$ & $1163.834\pm 0.023$ & $-12.822 \pm 0.459$ & {$0.574\pm 0.042$}  & {$0.724 \pm 0.329$} & {$1.112 \pm 0.102$} &  {$1.151 \pm 0.680$}\\

       N I  & $1168.536$ & $1168.478\pm 0.027$ & $-14.866 \pm 0.466$ & {$0.748\pm 0.064$}  & {$0.962 \pm 0.165$} & {$1.294 \pm 0.137$} &   {$0.703 \pm 0.157$}\\

        C III & $1174.93$ & {$1174.981 \pm 0.003$} & $13.105$ &  {$0.496 \pm 0.039$}&  {$6.164 \pm 3.725$} &{$0.395 \pm 0.015$} &   {$0.463 \pm 0.076$}\\

         C III & $1175.26$ & {$1175.368 \pm 0.005$} & $11.852$ &  {$0.158 \pm 0.050$}&  {$-$} &{$0.262 \pm 0.032$} &   {$0.424 \pm 0.307$}\\

        C III & $1175.59$ & {$1175.648 \pm 0.056$} & $14.814$ &  {$1.383 \pm 0.348$}&  {$-$} &{$1.444 \pm 0.136$} &   {$0.506 \pm 0.398$}\\

        C III & $1175.71$ & {$1175.748 \pm 0.003$} & $9.633$ &  {$1.195 \pm 0.217$}&  {$-$} &{$0.465
        \pm 0.031$} &   {$0.773 \pm 0.145$}\\

        C III & $1175.99$ & {$1176.110 \pm 0.010$} & $30.594$ &  {$0.174 \pm 0.038$}&  {$0.513 \pm 0.444$} &{$0.339
        \pm 0.038$} &   {$0.267 \pm 0.091$}\\

        C III& $1176.37$ & {$1176.420 \pm 0.006$} & $12.855$ &  {$0.330 \pm 0.042$}&  {$1.212 \pm 0.087$} &{$0.338
        \pm 0.022$} &   {$0.316 \pm 0.017$}\\

         C I & {$1190.42$} & {$1190.376 \pm 0.011$} & {$-10.979 \pm 0.242$} &  {$1.112 \pm 0.043$}&  {$2.248 \pm 0.241$} &{$0.873\pm 0.043$} &   {$0.687 \pm 0.096$}\\

        C I  & $1194.486 \pm 0.003$ & $1194.396\pm 0.012$ & $-23.582 \pm 0.142$ & {$2.202\pm 0.096$}  & {$5.315 \pm 0.881$} & {$1.087 \pm 0.060$} &   {$1.239 \pm 0.237$}\\
        
        Si II  & $1197.39$ & $1197.384\pm 0.010$ & $-1.607 \pm 1.493$ & {$0.899\pm 0.033$}  & {$1.475 \pm 0.167$} & {$0.762 \pm 0.034$} &   {$0.780 \pm 0.117$}\\

        N I  & $1199.55$ & $1199.654\pm 0.007$ & $26.011 \pm 0.067$ & {$21.692\pm 0.460$}  & {$45.206 \pm 4.807$} & {$1.567 \pm 0.039$} &   {$1.741 \pm 0.221$}\\

        O IV  & $1204.36$ & $1204.350\pm 0.013$ & $-2.448 \pm 1.280$ & {$1.375\pm 0.055$}  & {$2.471 \pm 0.862$} & {$1.127 \pm 0.057$} &   {$1.685 \pm 0.711$} & [2]\\

         Si III & $1206.51$ & $1206.630 \pm 0.014$ & $29.830 \pm 0.116 $ & {$15.201 \pm 0.358$}  & {$-$} & {$0.535 \pm 0.007$} &   {$-$}& [1] [3] \\

         O V] & $1218.344$ & $1218.223\pm 0.012$ & $-29.944 \pm 0.101$ & {$17.854 \pm 0.681$}  & {$24.698 \pm 2.572$} & {$1.355 \pm 0.063$} &   {$1.530 \pm 0.194$} & [4][5]\\

        N V   & {$1238.804 \pm 0.006$} & {$1238.831 \pm 0.001$} & {$6.650 \pm 0.223$} & {$10.274 \pm 0.010$} & {$1.302 \pm 0.408$} & {$0.432 \pm 0.005$} &  {$0.383 \pm 0.015$}\\

        Fe XII   & {$1242.00 \pm 0.02$} & {$-$} & {$-$} & {$-$} &  {$-$} &{$-$} &   {$-$}& [5]\\

        N V   & {$1242.795 \pm 0.006$} & {$1242.843 \pm 0.002$} & {$11.547 \pm 0.131$} & {$5.337 \pm 0.060$} &  {$6.747 \pm 0.326$} &{$0.453 \pm 0.006$} &   {$0.432 \pm 0.027$}\\

        C I & $1247.383$ & $1247.369 \pm 0.032$ & $-3.325 \pm 2.285$ & {$0.751 \pm 0.093$}  & {$-$} & {$0.955 \pm 0.107$} &   {$0.700 \pm 0.190$}& [6][7] \\

        C I  & $1248.003 \pm 0.003$ & $1248.146 \pm 0.029$ & $34.407 \pm 0.202$ & {$0.506 \pm 0.088$}  & {$-$} & {$0.756 \pm 0.106$} &   {$0.692 \pm 0.381$}& [6]\\

        C I  & $1249.404 \pm 0.003$ & $1249.278 \pm 0.033$ & $-30.121 \pm 0.262$ & {$0.651 \pm 0.067$}  & {$0.901 \pm 0.243$} & {$1.452 \pm 0.179$} &   {$0.964 \pm 0.335$}\\

        S II   & {$1250.58 \pm 0.002$} & {$1250.564 \pm 0.008$} & {$-3.742 \pm 0.543$} & {$0.632 \pm 0.023$} &  {$0.799 \pm 0.078$} &{$0.637\pm 0.028$} &   {$0.425 \pm 0.053$}\\

        Fe II   & {$1251.254 \pm 0.002$} & {$1251.181 \pm 0.021$} & {$-17.502 \pm 0.286$} & {$0.580 \pm 0.041$} &  {$0.729 \pm 0.114$} &{$1.020\pm 0.089$} &   {$0.581 \pm 0.117$}\\

        C I  & $1252.209 \pm 0.003$ & $1252.241 \pm 0.017$ & $7.702 \pm 0.524$ & {$0.382 \pm 0.024$}  & {$-$} & {$0.769 \pm 0.058$} &   {$-$}& [1] \\
        
        S II  & $1253.813 \pm 0.004$ & $1253.754 \pm 0.017$ & $-14.077 \pm 0.290$ & {$1.059 \pm 0.061$}  & {$1.697 \pm 0.245$} & {$1.079 \pm 0.079$} &   {$0.970 \pm 0.188$}\\

        C I  & $1254.51 \pm 0.003$ & $1254.017 \pm 0.399$ & $-117.726 \pm 0.809$ & {$1.374 \pm 0.684$}  & {$0.962 \pm 0.292$} & {$2.912 \pm 1.306$} &   {$1.010 \pm 0.345$}& [5]\\

        Si I  & {$1255.28$} & {$1255.296 \pm 0.026$} & {$3.780 \pm 1.664$} & {$0.481 \pm 0.039$} &  {$0.324 \pm 0.083$} &{$1.150\pm 0.116$} &   {$0.580 \pm 0.192$}\\

        C I  & {$1256.493 \pm 0.006 $} & {$1256.524 \pm 0.005$} & {$7.285 \pm 0.264$} & {$0.553 \pm 0.017$} &  {$0.786 \pm 0.141$} &{$0.502\pm 0.019$} &   {$0.795 \pm 0.189$}\\

        C I  & {$1257.565 \pm 0.003 $} & {$1257.593 \pm 0.012$} & {$6.786 \pm 0.443$} & {$0.400 \pm 0.019$} &  {$ 1.046\pm 0.523$} &{$0.735\pm 0.043$} &   {$1.582 \pm 0.891$}\\
        
     Si I  & {$1258.8$} & {$1258.827 \pm 0.008$} & {$6.323 \pm 0.296$} & {$0.487 \pm 0.024$} &  {$-$} &{$0.469 \pm 0.028$} &   {$-$}& [1][6]\\

     S II  & {$1259.516$} & {$1259.557 \pm 0.005$} & {$9.751 \pm 0.125$} & {$0.798 \pm 0.023$} &  {$-$} &{$0.533 \pm 0.018$} &   {$-$}& [1][6]\\

      C I  & $1260.612 \pm 0.003$ & $1260.559 \pm 0.016$ & $-12.717\pm 0.302$ & {$1.690 \pm 0.084$}  & {$2.538 \pm 0.339$} & {$1.230 \pm 0.076$} &   {$1.022 \pm 0.179$} \\

      C I  & $1261.551 \pm 0.003$ & $1261.357 \pm 0.055$ & $-46.153 \pm 0.285$ & {$1.540 \pm 0.239$}  & {$1.490 \pm 0.413$} & {$1.862 \pm 0.323$} &   {$1.107 \pm 0.382$}&  \\

        Si II   & $1264.73$ & {$1264.808 \pm 0.004$} & $18.528 \pm 0.056$ & {$2.819 \pm 0.056$} &  {$4.215 \pm 0.258$} &{$0.654 \pm 0.016$} &   {$0.531 \pm 0.042$}\\

        C I  & {$1266.52 \pm 0.003 $} & {$1266.443 \pm 0.007$} & {$-18.116 \pm 0.097$} & {$0.715 \pm .021$} &  {$0.696 \pm 0.084$} &{$0.692\pm 0.026$} &   {$0.438 \pm 0.067$}\\

         C I  & {$1267.599 \pm 0.003 $} & {$1267.619 \pm 0.011$} & {$4.722 \pm 0.586$} & {$0.378 \pm 0.020$} &  {$0.675 \pm 0.128$} &{$0.578\pm 0.036$} &   {$0.763 \pm 0.191$}\\

           S I & $1269.209 \pm 0.001$ & $1269.195 \pm 0.023$ & $-3.132 \pm 1.749$ & {$0.363 \pm 0.027$}  & {$0.530 \pm 0.138$} & {$1.173 \pm 0.110$} &   {$0.948 \pm 0.333$} \\

            Ni V  & $1270.69 \pm 0.009$ & $1270.781 \pm 0.035$ & $21.413 \pm 0.402$ & {$0.456 \pm 0.048$}  & {$0.358 \pm 0.314$} & {$1.236 \pm 0.156$} &   {$1.120 \pm 0.998$} \\

\end{tabular} 
   
    \label{tab:line_catalog}

\end{table}

\begin{table}
\centering
\small
\setlength{\tabcolsep}{1.5pt}
\begin{tabular}{l|l|l|l|l|l|l|l|c}

       Ion    & $\lambda_{\text{rest}} [\textnormal{\AA{}}]$  & $\lambda_{\text{obs}} [\textnormal{\AA{}}]$ & V$_{\text{Shift}} [\text{km} \, \text{s}^{-1}]$ & Flux (Q) & Flux (F) & FWHM (Q) &   FWHM (F) & Notes\\

        \hline \hline

        C I  & $1288.422 \pm 0.003$ & $1288.312 \pm 0.015$ & $-25.496 \pm 0.139$ & {$0.365 \pm 0.023$}  & {$1.152 \pm 0.115$} & {$0.688 \pm 0.052$} &   {$0.411 \pm 0.052$} & \\

        S IV   & $1294.604$ & {$1294.572 \pm 0.005$} & $-7.420 \pm 0.147$ & {$1.106 \pm 0.028$} &  {$3.985 \pm 0.235$} &{$0.515 \pm 0.016$} &   {$0.391 \pm 0.029$}&\\

         Si III  & $1296.73$ & $1296.794 \pm 0.011$ & $14.711 \pm 0.180$ & {$0.902 \pm 0.044$}  & {$2.226 \pm 0.243$} & {$0.647 \pm 0.039$} &   {$0.433 \pm 0.060$}&\\

        Si III   & $1298.96$ & {$1299.027 \pm 0.003$} & $15.448 \pm 0.045$ & {$2.506 \pm 0.043$} &  {$5.858 \pm 0.326$} &{$0.499 \pm 0.011$} &   {$0.560 \pm 0.040$}&\\

        S I   & {$1309.3 \pm 0.15$} & {$1309.316 \pm 0.003$} & {$-$}  & {$1.349\pm 0.027$} &  {$1.651 \pm 0.157$} &{$0.461 \pm 0.012$} &   {$0.270 \pm 0.032$}&\\
  
        C I   & {$1310.636 \pm 0.003$} & {$1310.640 \pm 0.012$} & {$-$}  & {$0.677 \pm 0.033$} & {$0.691 \pm 0.101$} &{$0.685 \pm 0.040$} &  {$0.285 \pm 0.053$}&\\

        C I   & {$1311.362 \pm 0.003$} & {$1311.370 \pm 0.008$} & {$1.762 \pm 1.148$} & {$0.677 \pm 0.031$} &  {$1.056 \pm 0.092$} & {$0.495 \pm 0.028$} &   {$0.305 \pm 0.034$}&\\

        C I  & $1311.925 \pm 0.003$ & $-$ & $-$ & {$-$}  & {$-$} & {$-$} &   {$-$}& [5] \\

        C I   & {$1313.462 \pm 0.003$} & {$1313.463 \pm 0.007$} & {$-$} & {$0.430 \pm 0.016$} &  {$0.923 \pm 0.191$} & {$0.503 \pm 0.023$} &  {$0.671 \pm 0.181$}&\\

        C I   & {$1315.918 \pm 0.003$} & {$1315.969 \pm 0.020$} & {$11.612 \pm 0.401$} & {$0.481 \pm 0.034$} & {$0.223 \pm 0.037$} & {$0.772 \pm 0.066$} &   {$0.070 \pm 0.015$}&\\

        S I   & {$1316.618 \pm 0.003$} & {$1316.584 \pm 0.009$} & {$-7.756 \pm 0.266$} & {$0.432 \pm 0.021$} &  {$0.365 \pm 0.072$} & {$0.509 \pm 0.030$} &   {$0.197 \pm 0.049$}&\\

        N I   & $1319$ & {$1318.985 \pm 0.007$} & $-3.415 \pm 0.495$ & {$0.415 \pm 0.018$} & {$0.387 \pm 0.076$} &{$0.467 \pm 0.025$} &  {$0.454 \pm 0.113$}&\\

         N I   & $1319.68$ & {$1319.635 \pm 0.029$} & $-10.320 \pm 0.638$ & {$0.472 \pm 0.042$} & {$-$} &{$1.240 \pm 0.137$} &  {$-$}& [1]\\

          S I   & $1323.522 \pm 0.0008$ & {$1323.722 \pm 0.036$} & $45.398 \pm 0.181$ & {$0.957 \pm 0.074$} & {$-$} &{$1.239 \pm 0.093$} &  {$-$}& [1][6]\\

         C II   & $1323.948$ & {$1324.04 \pm 0.009$} & $17.240 \pm 0.119$ & {$0.135 \pm 0.031$} & {$-$} &{$0.237 \pm 0.039$} &  {$-$}& [1][6] \\
        
        S I  & {$1326.643 \pm 0.001$} & {$1326.648 \pm 0.016$} & {$-$}  & {$0.436 \pm 0.027$}  & {$-$} & {$0.726\pm 0.055$} &   {$-$}& [1]\\

         C I   & $1328.83$ & {$1328.83$} & $0$ & {$0.390$} & {$0.343 \pm 0.068$} &{$0.235$} &  {$0.063 \pm 0.016$} & [8]\\

         C I   & $1329.123$ & {$1329.123$} & $0$ & {$0.427$} & {$0.780 \pm 0.132$} &{$0.235$} &  {$0.277 \pm 0.062$} & [8]\\

         C I & $1329.600$ & {$1329.600$} & $0$ & {$0.400$} & {$0.746 \pm 0.104$} &{$0.235$} &  {$0.199 \pm 0.035$} & [1][8]\\

         C II   & $1334.532 \pm 0.0005$ & {$1334.643 \pm 0.006$} & $24.865 \pm 0.052$ & {$17.314 \pm 0.535 $} & {$22.259 \pm 0.893$} &{$0.515 \pm 0.020$} &  {$0.421 \pm 0.022$} & [3]\\

        C II   & {$1335.708 \pm 0.001$} & {$1335.729 \pm 0.001$} & {$4.697\pm 0.062$} & {$35.525 \pm 0.269$} &  {$49.242 \pm 1.039$} & {$0.477 \pm 0.005$} &   {$0.449 \pm 0.012$}&\\
        
        Cl I   & $1351.657 $ & {$1351.674 \pm 0.003$} & $3.705 \pm 0.150$ & {$1.399 \pm 0.035$} &  {$1.661 \pm 0.076$} & {$0.280 \pm 0.009$} &   {$0.161 \pm0.009$}&\\

       Fe XXI  & {$1354.08\pm 0.002$} & {$1354.085\pm 0.004$} & {$1.033 \pm 0.998$} & {$4.119 \pm 0.067$} &  {$4.630\pm 0.313$} & {$0.740 \pm 0.015$} &  {$0.682 \pm 0.060$}&\\

        C I   & {$1357.134 \pm 0.003$} & {$1357.159 \pm 0.013$} & {$5.534 \pm 0.530$} & {$0.658 \pm 0.034$} & $-$ & {$0.698 \pm 0.044$} &  $-$& [1]\\

        Fe V   & $1358.567 \pm 0.004$ & {$1358.485 \pm 0.010$} & $-18.109 \pm 0.134$ & {$1.034 \pm 0.041$} & {$0.932 \pm 0.125$} &{$0.758 \pm 0.038$} &  {$0.326 \pm 0.055$}&\\

       C I   & {$1359.276 \pm 0.003$} & {$1359.241 \pm 0.019$} & {$-7.721 \pm 0.547$} & {$0.774 \pm 0.050$} &  {$1.006 \pm 0.262$} & {$0.954 \pm 0.078$} &   {$0.914 \pm 0.319$}&\\

       Fe II   & {$1361.359 \pm 0.002$} & {$1361.402 \pm 0.007$} & {$9.481 \pm 0.168$} & {$0.572 \pm 0.024$} & {$0.590 \pm  0.074$} & {$0.444 \pm 0.023$} &  {$0.204 \pm 0.032$}&\\

       C I   & {$1364.164 \pm 0.003$} & {$1364.186 \pm 0.008$} & {$4.815 \pm 0.377$} &{$0.557 \pm 0.024$} &  {$1.076 \pm 0.149$} & {$0.480 \pm 0.025$} &   {$0.559 \pm 0.099$}&\\

       O V   & $1371.292$ & {$1371.317 \pm 0.005$} & $5.481 \pm 0.195$ & {$1.138 \pm 0.028$} & {$1.511 \pm 0.190$} & {$0.544 \pm 0.017$} &  {$0.536 \pm 0.086$}&\\
       
       Si IV   & $1393.76$ & {$1393.787\pm 0.001$} & $5.817 \pm 0.055$ & {$24.975 \pm 0.203$} &  {$49.238 \pm 0.890$} &{$0.518 \pm 0.005$} &   {$0.497 \pm 0.011$}&\\

       O IV]   & $1399.774 $& {$1399.812 \pm 0.011$} & $8.171 \pm 0.282$ & {$0.906 \pm 0.034$} & {$0.737 \pm 0.107$} &{$0.885 \pm 0.042$} &  {$0.465 \pm 0.086$}& [7] \\

       O IV]   & $1401.156 $ & {$1401.214 \pm 0.007$} & $12.407 \pm 0.128$ & {$1.826 \pm 0.053$} & {$2.646 \pm 0.595$} &{$0.736 \pm 0.027$} &  {$1.001 \pm 0.295$}& [7]\\

        Si IV   & $1402.77$ & {$1402.811 \pm 0.002$} & $8.698 \pm 0.038$ & $13.311 \pm 0.111$ &  {$29.545\pm 0.703$} & {$0.518 \pm 0.005$} &   {$0.523 \pm 0.016$}&\\

        S IV]+  & {$1404.77 /0.812$} & {$1404.781 \pm 0.015$} & {$-$} & {$1.033 \pm 0.052$} &  {$0.813 \pm 0.234$} &{$1.055\pm 0.067$} &   {$0.744 \pm 0.282$}& [4]\\

        \text{  O IV}] &&&&&&&& \\

        S IV]   & $1406.009$ & {$1406.048 \pm 0.017$} & $8.243 \pm 0.428$ & {$0.871 \pm 0.048$} & {$0.904 \pm 0.152$} &{$0.989 \pm 0.068$} &  {$0.701 \pm 0.155$} &\\

         O IV]  & {$1407.361$} & {$1407.391 \pm 0.015$} & {$6.405 \pm 0.505$} & {$0.918 \pm 0.045$} &  {$0.530 \pm 0.095$} &{$1.078\pm 0.067$} &   {$0.223 \pm 0.051$}& [4]\\

        N I   &$1411.94$ & {$1411.938 \pm 0.015$} & $-$ & {$0.756 \pm 0.042$} &  {$0.943 \pm 0.186$} & {$0.772 \pm 0.052$} &   {$0.712 \pm 0.184$}&\\

        Fe III   & {$1413.774 \pm 0.005$} & {$1413.720 \pm 0.016$}& {$ -11.428 \pm 0.317$} & {$0.709 \pm 0.039$} &  {$0.679 \pm 0.254$} & {$0.917 \pm 0.062$} & {$1.000 \pm 0.500$}&\\

       Fe V   & {$1425.088 \pm 0.004$} & {$1425.109 \pm 0.007$} & {$4.484 \pm 0.398$} & {$1.028 \pm 0.036$}  & {$1.166 \pm 0.228$} & {$0.597 \pm 0.026$} &   {$0.809 \pm 0.210$} &\\

      Fe V   & $1426.46 \pm 0.005$ & {$1426.371 \pm 0.023$} & $-18.664 \pm 0.260$ & {$0.783 \pm 0.060$} & {$0.659 \pm 0.067$} &{$1.162 \pm 0.112$} &  {$0.222 \pm 0.029$}& \\

\end{tabular}

\caption{All flux values are in units of $10^{-15}$ erg s$^{-1}$ cm$^{-2}$. All FWHM values are in units of \AA. Q refers to quiescent flux, F refers to flaring flux over flare 18. Emission lines due to airglow are not included in this line list. Rest wavelengths without a listed error are approximated as 0. 
[1] = Flare 18 components could not be resolved. 
[2] = Line identified using Ritz wavelength in NIST \cite{NIST_ASD}
[3] = Absorption feature identified in line
[4] = Rest wavelength from \cite{pagano_paper}
[5] = Line identified but features unresolved.
[6]= Fit with double pseudovoigt profile 
[7] = Rest wavelength from \cite{feinstein_paper}
[8] = Fit with a triple pseudovoigt profile with completely constrained centroid. }

\label{tab:line_catalog}
    
\end{table}
\clearpage
\bibliography{ds_tuc_ref}{}
\bibliographystyle{aasjournalv7}

\end{document}